\documentclass{aa}  

\usepackage{graphicx}
\usepackage{txfonts}
\usepackage{xcolor}
\usepackage{lipsum}
\definecolor{mygray}{gray}{0.9}
\begin{document}

   \title{Horizontal shear instabilities in rotating stellar radiation zones }

   \subtitle{I. Inflectional and inertial instabilities and the effects of thermal diffusion}

   \author{J. Park,
          V. Prat
          %\inst{1}
          \and
          S. Mathis
          %\inst{1}
          }

   \institute{AIM, CEA, CNRS, Universit\'e Paris-Saclay, Universit\'e Paris Diderot, Sorbonne Paris Cit\'e, F-91191 Gif-sur-Yvette, France\\
              \email{junho.park@cea.fr}
                           }

   \date{}

% \abstract{}{}{}{}{} 
% 5 {} token are mandatory
 
  \abstract
   {{Rotational mixing transports} angular momentum and chemical elements in stellar radiative zones. 
   It is one of the key processes for modern stellar evolution. 
   {In the past two decades, {an} emphasis has been placed} on the turbulent transport induced by the vertical shear instability. 
   However, instabilities arising from horizontal shear and the strength of the anisotropic turbulent transport that they may trigger remain relatively unexplored.
   {The weakest point of this hydrodynamical theory of rotational mixing is the assumption} that anisotropic turbulent transport is stronger in horizontal directions than in the vertical one. 
   }
   %The so-called rotational mixing, which transports angular momentum and chemical elements in stellar radiative zones, is one of the key processes for modern stellar evolution. In the two last decades, the stress has been put on the turbulent transport induced by the vertical shear instability. However, the instabilities of horizontal shears and the strength of the anisotropic turbulent transport they may trigger are still largely unknown. This is the weakest point of the current theory of the hydrodynamical rotational mixing in which we assume an anisotropic turbulent transport stronger along the horizontal directions than along the vertical one.}}
   {{This paper investigates} the combined effects of stable stratification, rotation, and thermal diffusion on the horizontal shear instabilities {that are obtained and} discussed in the context of stellar radiative zones.
   %This paper investigates for the first time the combined effects of stable stratification, rotation, and thermal diffusion on the instabilities of horizontal shears. The obtained general results are discussed in the context of stellar radiative zones.
   %on two kinds of horizontal shear flow instabilities found in stably stratified, rotating flows: the inflectional and inertial instabilities.
   }
   {The eigenvalue problem describing linear instabilities of a flow with a hyperbolic-tangent horizontal shear profile was solved numerically for a wide range of parameters.
   {When possible,} the Wentzel-Kramers-Brillouin-Jeffreys (WKBJ) approximation was applied to provide {analytical} asymptotic dispersion relations in {both} the nondiffusive and highly diffusive limits.  
   As a first step, we consider a polar $f$-plane where the gravity and rotation vector are aligned.
   
   }
   %The eigenvalue problem describing the linear instabilities of a flow with a hyperbolic-tangent horizontal shear profile is solved numerically for a wide range of parameters.
   %Furthermore, the Wentzel-Kramers-Brillouin-Jeffreys (WKBJ) approximation is applied when this is possible to provide explicit asymptotic dispersion relations in non-diffusive and highly diffusive limits. As a first step, we consider a polar f-plane where the gravity and the rotation vector are aligned.}
   {
   Two types of instabilities are identified: the inflectional and inertial instabilities. 
   {The inflectional instability that arises from the inflection point (i.e., the zero second derivative of the shear flow)} is the most unstable when at a zero vertical wavenumber and a finite wavenumber in the streamwise direction along the imposed-flow direction.
   While the maximum two-dimensional growth rate is independent of the stratification, rotation rate, and thermal diffusivity, the three-dimensional inflectional instability is destabilized by stable stratification, while it is stabilized by thermal diffusion.
   {The inertial instability is rotationally driven, and a WKBJ analysis reveals that} its growth rate reaches the maximum value of $\sqrt{f(1-f)}$ {in the inviscid limit as the vertical wavenumber goes to infinity, where $f$ is the dimensionless Coriolis parameter}. 
   The inertial instability for a finite vertical wavenumber is stabilized as the stratification increases, {whereas it is destabilized by the thermal diffusion.
   Furthermore, we found a selfsimilarity in both the inflectional and inertial} instabilities based on the rescaled parameter $PeN^2$ with the P\'{e}clet number $Pe$ and the Brunt-V\"{a}is\"{a}l\"{a} frequency $N$.
   
    }
   {}


   \maketitle
%
%-------------------------------------------------------------------

\section{Introduction}

%Transport processes in stellar interiors are crucial for stellar structure and evolution because they redistribute angular momentum and chemical elements.
%In particular, rotational mixing, which refers to transport processes associated with stellar rotation, is a key ingredient of stellar evolution.
%\citet{Zahn1992} initially proposed a formalism based on the interaction of advection by the meridional circulation and diffusion by shear instabilities.
%The author used phenomenological arguments to derive analytical expressions for vertical and horizontal diffusion coefficients associated with these instabilities.

%In the last decades, more sophisticated transport coefficients have been proposed, mostly in the vertical direction \citep{Maeder1995, MaederMeynet1996, Maeder1997, TalonZahn1997}, and more rarely in the horizontal direction \citep{Maeder2003, Mathis2004}.

%The development of helio- and asteroseismology has provided us with new constraints on transport processes.
%Many of those constraints suggest that current models of rotational mixing do not predict enough transport of angular momentum.
   
%  \vec{Astrophysical introduction 2: Why horizontal shear instability in stratified-rotating fluids is important in astrophysics? } \textcolor{mygray}{\lipsum[3]}

The combination of space-based helio- and asteroseismology has demonstrated that stably stratified rotating stellar radiation zones are the seats of efficient transport of angular momentum {throughout} the evolution of stars. 
This strong transport leads to a uniform rotation in the case of the Sun down to $0.2R_{\odot}$ \citep{Garciaetal2007} and to weak differential rotation in other stars \citep[e.g.,][]{Mosseretal2012,Deheuvelsetal2012,Deheuvelsetal2014,Kurtzetal2014,Saioetal2015,Murphyetal2016,Spadaetal2016,VanReethetal2016,Aertsetal2017,VanReethetal2018,Gehanetal2018,Ouazzanietal2019}. 
Four main mechanisms that transport angular momentum and mix chemicals are present in stellar radiation zones \citep[e.g.,][and references therein]{Maeder2009,Mathis2013,Aertsetal2019}: instabilities of the differential rotation \citep[e.g.,][]{Zahn1983,Zahn1992}, stable and unstable magnetic fields \citep[e.g.,][]{Spruit1999,Fulleretal2019}, internal gravity waves \citep[e.g.,][]{ Zahnetal1997,TalonCharbonnel2005,Pinconetal2017}, and large-scale meridional circulations \citep [e.g.,][]{Zahn1992,MaederZahn1998,MathisZahn2004}. 
Important {progress has} been made in their {modeling during the last two decades. 
However, many simplifying assumptions are still employed in the treatment of these four mechanisms} because of their complexity and of the broad range of spatial and temporal scales they involve. 
For instance, the complex interplay between rotation and stratification for the study of vertical and horizontal shear instabilities is partially treated. 
On the one hand, the action of the Coriolis acceleration is not taken into account in the state-of-the-art modeling of the vertical turbulent transport due to the instabilities {by} radial differential rotation. 
On the other hand, the important action of thermal diffusion has completely been ignored in the studies of the horizontal turbulent transport induced by {horizontal and vertical shear instabilities \citep[see e.g.,][respectively]{Mathisetal2004,Mathisetal2018}. 
As a consequence, a lot of work is still required to obtain} robust abinitio prescriptions for transport processes in stars and to reconcile stellar models and the observations \citep[e.g.,][]{Eggenbergeretal2012,Ceillieretal2013,Marquesetal2013,Cantielloetal2014,Eggenbergeretal2019}. 
In this work, our aim is to improve our understanding of horizontal shear instabilities in rotating stellar radiation zones.    

In his seminal article, \cite{Zahn1992} built the theoretical framework to study the turbulent transport induced by vertical and horizontal shears in convectively stable rotating stellar radiation zones. 
In such regions, turbulence can be anisotropic because of the combined action of buoyancy and the Coriolis acceleration, which control the turbulent motions along the vertical and the horizontal directions, respectively \citep[e.g.,][]{BillantChomaz2001,Davidson2013,Mathisetal2018}. 
As pointed out in \cite{Mathisetal2018}, many theoretical works have been devoted to {providing} robust prescriptions for the vertical eddy diffusion associated with the instability of vertical shear \citep[e.g.,][]{Zahn1992,Maeder1995,MaederMeynet1996,Maeder1997,TalonZahn1997, KulenthirarajahGaraud2018}. 
Their predictions are now {being} tested using direct numerical simulations \citep[][]{PratLignieres2013, PratLignieres2014,Pratetal2016,Garaudetal2017,GagnierGaraud2018}. 
However, very few studies have examined {the instabilities arising from horizontal shear} \cite[e.g.,][]{Zahn1992,Maeder2003,Mathisetal2004}. 
The resulting prescriptions are based mostly on phenomenological approaches and {the results of unstratified Taylor-Couette flow experiments} \citep[][]{RichardZahn1999} that do not take into account the complex interplay of stratification, rotation, and thermal diffusion, {all of which play important roles} in stellar radiation zones. 
For instance, heat diffusion inhibits the effects of the stable stratification for the vertical shear instability \citep{Townsend1958,Zahn1983}. 
\cite{Mathisetal2018} studies the combined action of stratification and rotation on the horizontal turbulent transport. 
{Their approaches to the turbulent transport have been successfully implemented in several stellar evolution codes \citep[][]{Talonetal1997,MeynetMaeder2000,Palaciosetal2006,Marquesetal2013}. 
However, the neglect of the nonadiabatic aspects of the problem and the focus only on the latitudinal turbulent transport induced by the three-dimensional motions resulting from the vertical shear instability constitute weak points of the theory for the rotational mixing.
%But they have neglected the non-adiabatic aspects of the problem and they only focused on the latitudinal turbulent transport induced by the three-dimensional motions resulting from the instability of a vertical shear. 
%This constitutes an important weak point of the theory for the rotational mixing, which has been successfully implemented in several stellar evolution codes \citep[][]{Talonetal1997,MeynetMaeder2000,Palaciosetal2006,Marquesetal2013}. 
Indeed, in the formalism of \cite{Mathisetal2018}, a stronger turbulent transport is assumed along the horizontal direction than along the vertical direction}. 
This allows us to assume that the horizontal gradients of the angular velocity (and of the fluctuations of temperature and chemical composition) are smoothed out, leading to the so-called “shellular” rotation, that only varies with the radius. 
Such a radial variation of rotation is pertinent to the vertical shear instability, which has been mostly studied on the local $f$-plane with vertical stratification \citep[see e.g.,][]{Wang2014}.  
The horizontal shear instability with stellar differential rotation in latitudinal direction has also been studied \citep[see e.g.,][]{Watson1980,Garaud2001,Kitchatinov2009}, but the combined impacts of the rotation, stratification, and thermal diffusion on horizontal shear instability are not yet fully resolved.
It is thus mandatory to study the instabilities of horizontal shear in rotating and stably stratified stellar regions (Fig.~\ref{Fig_cartoon}a,b) while taking into account their important thermal diffusion.

In classical fluid dynamics, theories of shear flow stability in homogeneous fluids are well {known} \citep[]{Schmid2001}. 
According to Rayleigh's inflection point criterion, an inviscid unstable shear flow should possess an inflectional point where the second derivative of the zonal flow velocity $U$ vanishes (i.e., $U''(y)=0$, where $y$ is the local latitudinal coordinate; see Fig.~\ref{Fig_cartoon}c). 
For instance, a shear flow with a hyperbolic tangent velocity profile can undergo this inflection point instability \citep[]{Michalke1964}. 
In geophysical fluid dynamics, this instability is also referred {to} the barotropic instability \citep[]{Kundu2001} as it can be triggered by a two-dimensional horizontal perturbation with a finite streamwise zonal wavenumber $k_{\mathrm{x}}$. 
However, general three-dimensional perturbations with both $k_{\mathrm{x}}$ and a vertical wavenumber $k_{\mathrm{z}}$ can also trigger the inflectional instability of the horizontal shear flow.
So, we use the terminology inflectional instability hereafter. 
The stability analysis of such shear flow in stratified fluids by \citet{Deloncle2007} revealed that the unstable regime in the parameter space of zonal and vertical wavenumbers $(k_{\mathrm{x}},k_{\mathrm{z}})$ is widely broadened for a horizontal shear flow $U(y)$ in strongly stratified fluids.
This implies that perturbations with a smaller vertical scale (i.e., large $k_{\mathrm{z}}$) can still be unstable with a large Brunt-V\"{a}is\"{a}l\"{a} frequency $N$. 
A selfsimilarity of the three-dimensional inflectional instability is found for strong stratification with a rescaled parameter $Nk_{\mathrm{z}}$ \citep[]{Deloncle2007}. 
However, they found that the most unstable perturbation is still two-dimensional with a finite $k_{\mathrm{x}}$ at $k_{\mathrm{z}}=0$ independent of the stratification. 

The effects of both the stratification and rotation on horizontal shear instability were studied by \citet{Arobone2012}. 
They explored how the instability growth rate in the parameter space $(k_{\mathrm{x}},k_{\mathrm{z}})$ is modified as the stratification and rotation change. 
And they found that the most unstable mode of the inflectional instability is always found at finite $k_{\mathrm{x}}$ and $k_{\mathrm{z}}=0$ independent of the stratification and rotation. 
Moreover, they also underlined that there exists an inertial instability in a finite range of $f_{0}$ which is the Coriolis parameter defined as $f_{0}=2\Omega\cos\theta_{\rm s}$ where $\Omega$ is the rotation of a star and $\theta_{\rm s}$ is the colatitude. 
The origin of the inertial instability is different from the inflectional instability: the horizontal flow can become inertially unstable only in rotating fluids if the Rayleigh discriminant $\Phi(y)=f_{0}(f_{0}-U')$ becomes negative (Fig.~\ref{Fig_cartoon}d).
This condition leads to the inertially-unstable range $0<f_{0}<\max(U')$ for the hyperbolic tangent shear flow whose the maximum growth rate is found in inviscid limit as $\sqrt{f_{0}(\max(U')-f_{0})}$. 
{Such a mechanism is essentially equivalent to that of the centrifugal instability for rotating flows with cylindrical geometry. 
And it is centrifugally stable if the cylindrical Rayleigh discriminant $\Phi_{r}$ is always positive: 
\begin{equation}
\Phi_{r}=\frac{1}{r^{3}}\frac{\partial(ru_{\theta})^{2}}{\partial r}>0,
\end{equation}
where $u_{\theta}$ is the azimuthal velocity and $r$ is the cylindrical radial coordinate \citep[]{Kloosterziel1991,Park2013PoF}.
The stability condition can be extended by taking the stratification of the entropy $S$ into account.
This results in the Solberg-H\o iland conditions:
\begin{equation}
\begin{aligned}
&\Phi_{r}-\frac{1}{C_{\rm{p}}\rho}\nabla p\cdot\nabla S >0,\\
&\frac{\partial p}{\partial z}\left(\frac{\partial(ru_{\theta})^{2}}{\partial r}\frac{\partial S}{\partial z}-\frac{\partial S}{\partial r}\frac{\partial(ru_{\theta})^{2}}{\partial z}\right)<0,
\end{aligned}
\end{equation}
\citep[see also][]{Rudiger2002}.
Moreover, the Goldreich-Schubert-Fricke (GSF) criterion proposes the stability condition which takes effects of the thermal diffusivity $\kappa_{0}$ and viscosity $\nu_{0}$ into account:
\begin{equation}
\Phi_{r}+\frac{\nu_{0}}{\kappa_{0}}N^{2}>0,
\end{equation}
where $N$ is the Brunt-V\"ais\"al\"a frequency \citep[]{Goldreich1967,Fricke1968,Maeder2013}.
While these studies have proposed stability conditions in the presence of the stratification, rotation, and thermal diffusion, it is still not fully understood how horizontal shear instabilities are modified by the thermal diffusion, which has an essential importance for the dynamics of stellar radiative zones.}

In this paper, we investigate the linear stability of horizontal shear flows in stably-stratified, rotating, and thermally-diffusive fluids in the context of stellar radiation zones. In Sect.~2, equations for the linear stability analysis are formulated. 
In Sect.~3, we compute numerical results of this analysis for both the inflectional and inertial instabilities to {characterize} their main properties. 
In Sect.~4, these are compared with results from asymptotic analyses by means of the Wentzel-Kramers-Brillouin-Jeffreys (WKBJ) approximation for the inertial instability in order to understand the important role of thermal diffusion in stably-stratified rotating fluids. 
In Sect.~5, key scaling laws are derived for both the inflectional and inertial instabilities as a function of the stratification and thermal diffusivity. 
Finally, conclusions and perspectives of this work are presented in Sect.~6.

%--------------------------------------------------------------------
\section{Problem formulation}
\subsection{Governing equations and base equilibrium state}
%
%                                                One column figure
%----------------------------------------------------------------- 
   \begin{figure}
   \centering
   \includegraphics[height=8cm]{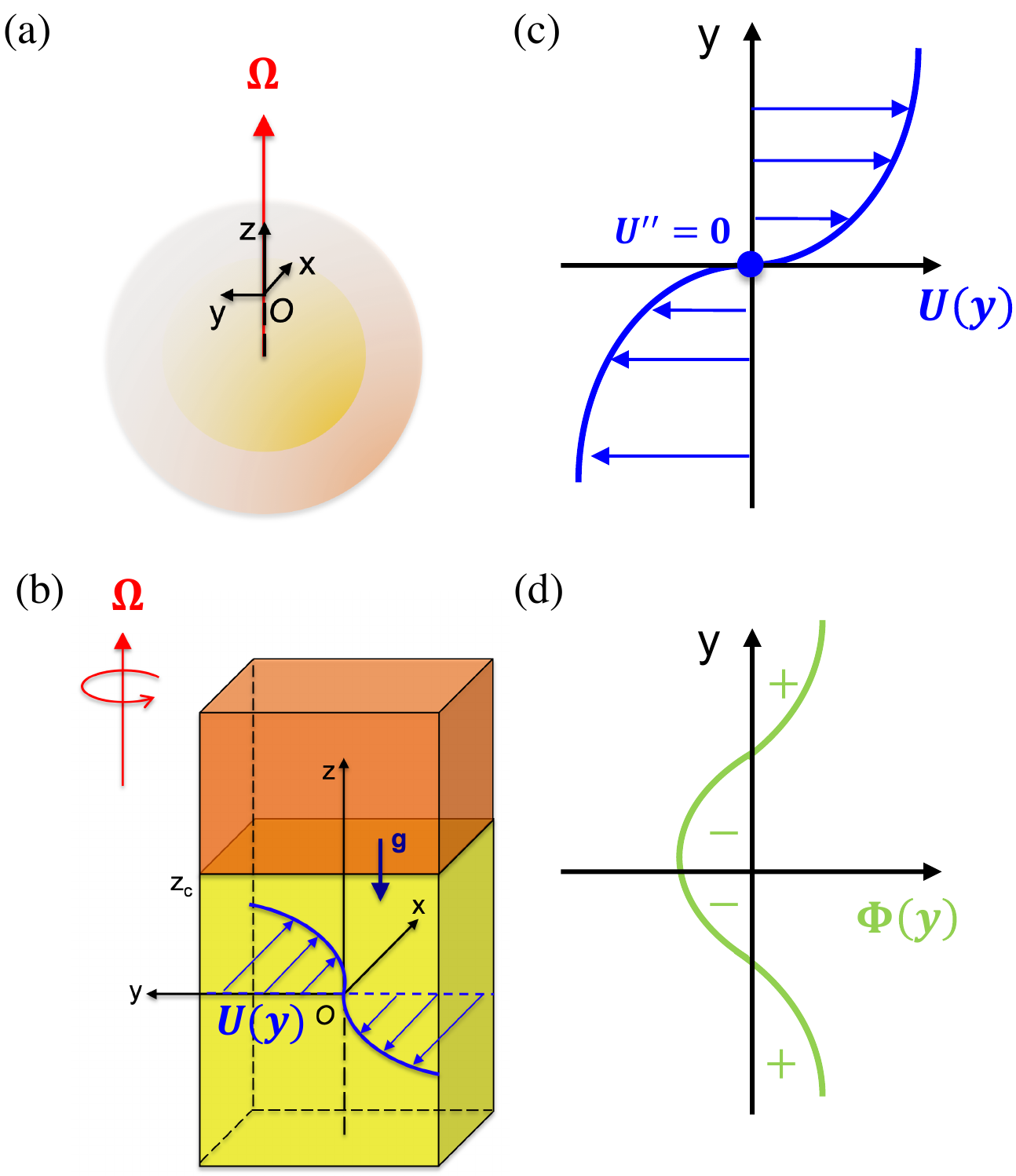}
      \caption{(a,b) Horizontal shear flow $U(y)$ on a local polar plane rotating with angular speed $\Omega$ in the radiative zone of a rotating star. The radiative and convective zones, colored as yellow and orange with the transition altitude $z_{c}$, are the configuration of low-mass stars; it should be inverted for early-type stars. (c) Horizontal shear flow profile $U(y)$ with an inflection point $U''=0$ for the inflectional instability. (d) Rayleigh discriminant $\Phi(y)=f_{0}\left(f_{0}-U'\right)$ for the inertial instability configuration.
      %Horizontal shear flow $U(y)$ on a local $f$-plane. The radiative and convective zones are colored as yellow and orange and $z_{c}$ is the transition altitude between the two zones, the configuration of low-mass stars. For the case of intermediate-mass and massive stars, the vertical structure is inverted. 
              }
         \label{Fig_cartoon}
   \end{figure}
%-----------------------------------------------------------------
We consider the Euler equations under the Boussinesq approximation and the heat transport equation in the Cartesian coordinate $(x,y,z)$ in a local frame rotating with angular velocity $\Omega$:
\begin{equation}
\label{eq:total_continuity}
	\nabla\cdot\vec{u}=0,
\end{equation}
\begin{equation}
\label{eq:total_momentum}
	\frac{\partial\vec{u}}{\partial t}+\left(\vec{u}\cdot\nabla\right)\vec{u}+\vec{f}\times\vec{u}=-\frac{1}{\rho_{0}}\nabla p-\alpha_{\rm T}\theta\vec{g},
\end{equation}
\begin{equation}
\label{eq:total_diffusion}
	\frac{\partial \theta}{\partial t}+\vec{u}\cdot\nabla \theta=\kappa_{0}\nabla^{2}\theta,
\end{equation}
where $\vec{u}=\left(u,v,w\right)$ is the velocity, $p$ is the pressure, $\theta=T-T_{0}$ is the temperature deviation from the reference temperature $T_{0}$, $\vec{f}=\left(0,0,2\Omega\right)$ is the Coriolis vector which is here antialigned with the gravity $\vec{g}=(0,0,-g)$, $\rho_{0}$ is the reference density, $\kappa_{0}$ is the reference thermal diffusivity, $\alpha_{\rm T}$ is the thermal expansion coefficient, and $\nabla^{2}$ denotes the Laplacian operator. We thus consider as a first step the traditional $f$-plane case corresponding to the horizontal shear flow on the pole (Fig.~\ref{Fig_cartoon}) to compare with previous literature. In this setup, the action of the latitudinal component of the rotation vector is filtered out \citep{Gerkema2005,Gerkema2008}. The current Cartesian coordinates $(x,y,z)$ on the traditional $f$-plane is associated with the spherical coordinates $(r,\theta_{\mathrm{s}},\varphi)$ where $r$ is the radial coordinate and $\varphi$ is the longitude. For instance, $x$ is the longitudinal coordinate with $\vec{e}_{\mathrm{x}}=\vec{e}_{{\varphi}}$ where $\vec{e}$ denotes the unit vector of the coordinate systems, $y$ is the latitudinal coordinate with $\vec{e}_{\mathrm{y}}=-\vec{e}_{{\theta_{\mathrm{s}}}}$, and $z$ is the vertical coordinate with $\vec{e}_{\mathrm{z}}=\vec{e}_{\mathrm{r}}$. 

To perform the linear stability analysis, we consider a steady base velocity $\vec{U}=\left(U(y),0,0\right)$ in a hyperbolic tangent form
\begin{equation}
\label{eq:base_shear}
	U=U_{0}\tanh\left(\frac{y}{L_{0}}\right),
\end{equation}
where $U_{0}$ and $L_{0}$ are the reference velocity and length scale, respectively. 
{Such a base flow is balanced with the pressure gradient as
\begin{equation}
\label{eq:base_pressure}
fU=-\frac{\partial\bar{P}}{\partial y},
\end{equation}
where $\bar{P}$ is the base pressure profile.
}
We consider a base temperature profile $\bar{\Theta}(z)$ which increases linearly with height $z$:
\begin{equation}
\label{eq:base_temperature}
	\bar{\Theta}=\frac{\Delta\Theta_{0}}{\Delta z}z,
\end{equation}
where $\Delta\Theta_{0}$ is the base-temperature difference along the vertical distance $\Delta z$.

\subsection{Linearized stability equations}
Subject to the base state, we consider perturbations $\tilde{\vec{u}}=\vec{u}-\vec{U}=\left(\tilde{u},\tilde{v},\tilde{w}\right)$, $\tilde{p}=p-\bar{P}$, and $\tilde{T}=\theta-\bar{\Theta}$. Hereafter, we use dimensionless parameters by converting the equations (\ref{eq:total_continuity}-\ref{eq:total_diffusion}) into a set of dimensionless equations with the length scale as $L_{0}$, the velocity scale as $U_{0}$, the time scale as $L_{0}/U_{0}$, the pressure scale as $\rho_{0}U_{0}^{2}$, and the temperature scale as $(L_{0}\Delta\Theta_{0})/\Delta z$. For infinitesimally small amplitude perturbations, we obtain the following linearized perturbation equations
\begin{equation}
\label{eq:ptb_continuity}
	\frac{\partial \tilde{u}}{\partial x}+\frac{\partial \tilde{v}}{\partial y}+\frac{\partial \tilde{w}}{\partial z}=0,
\end{equation}
\begin{equation}
\label{eq:ptb_x_mom}
	\frac{\partial \tilde{u}}{\partial t}+U\frac{\partial \tilde{u}}{\partial x}+\left(U'-f\right)\tilde{v}=-\frac{\partial \tilde{p}}{\partial x},
\end{equation}
\begin{equation}
\label{eq:ptb_y_mom}
	\frac{\partial \tilde{v}}{\partial t}+U\frac{\partial \tilde{v}}{\partial x}+f\tilde{u}=-\frac{\partial \tilde{p}}{\partial y},
\end{equation}
\begin{equation}
\label{eq:ptb_z_mom}
	\frac{\partial \tilde{w}}{\partial t}+U\frac{\partial \tilde{w}}{\partial x}=-\frac{\partial \tilde{p}}{\partial z}+N^{2}\tilde{T},
\end{equation}
\begin{equation}
\label{eq:ptb_diffusion}
	\frac{\partial \tilde{T}}{\partial t}+U\frac{\partial \tilde{T}}{\partial x}+\tilde{w}=\frac{1}{Pe}\nabla^{2}\tilde{T},
\end{equation}
where prime denotes the total derivative with respect to $y$, $N$ is the dimensionless Brunt-V\"ais\"al\"a frequency where
\begin{equation}
N^{2}=\frac{\alpha_{T}gL_{0}^{2}}{U_{0}^{2}}\frac{\Delta\Theta_{0}}{\Delta z},
\end{equation}
$f$ is the dimensionless Coriolis parameter 
\begin{equation}
f=\frac{2\Omega L_{0}}{U_{0}},
\end{equation}
and $Pe$ is the P\'eclet number
\begin{equation}
Pe=\frac{U_{0}L_{0}}{\kappa_{0}}.
\end{equation} 
We note that {$N$ is equivalent to the inverse of the horizontal Froude number $F_{h}$ for the stratified horizontal shear flow in \cite{Deloncle2007}. 
Also, $N^{2}$} is similar to the Richardson number $Ri$ defined as $Ri=N^{2}_{\mathrm{v}}/S^{2}_{\mathrm{v}}$ where $S_{\mathrm{v}}$ is the vertical shear and $N_{\mathrm{v}}$ is the Brunt-V\"ais\"al\"a frequency used for the vertical shear instability study \citep[]{Lignieresetal1999}. 
The nondimensional $f$ is equal to the inverse of the Rossby number $Ro=1/f$.

To perform a linear stability analysis, we consider a normal mode representation for the perturbation variables:
\begin{equation}
\label{eq:ptb_normal_mode}
	\left(\tilde{\vec{u}},\tilde{p},\tilde{T}\right)=\Re\left[\left(\hat{\vec{u}}(y),\hat{p}(y),\hat{T}(y)\right)\exp\left[\mathrm{i}\left(k_{\mathrm{x}} x+k_{\mathrm{z}} z\right)+\sigma t\right]\right],
\end{equation}
where $\mathrm{i}^{2}=-1$, $\hat{\vec{u}}=\left(\hat{u},\hat{v},\hat{w}\right)$, $\hat{p}$, and $\hat{T}$ are the latitudinal mode shapes for velocity, pressure, and temperature perturbation, respectively, $k_{\mathrm{x}}$ is the horizontal wavenumber in streamwise direction, $k_{\mathrm{z}}$ is the vertical wavenumber, and $\sigma=\sigma_{r}+\mathrm{i}\sigma_{i}$ is the complex growth rate where the real part $\sigma_{r}$ is the growth rate and the imaginary part $\sigma_{i}$ is the temporal frequency. 
{Expanding the equations (\ref{eq:ptb_continuity}-\ref{eq:ptb_diffusion}) with these normal modes}, we obtain the following linear stability equations
\begin{equation}
\label{eq:lse_continuity}
	\mathrm{i}k_{\mathrm{x}}\hat{u}+\frac{\partial\hat{v}}{\partial y}+\mathrm{i}k_{\mathrm{z}}\hat{w}=0,
\end{equation}
\begin{equation}
\label{eq:lse_x_mom}
	\left(\sigma+\mathrm{i}k_{\mathrm{x}} U\right)\hat{u}+\left(U'-f\right)\hat{v}=-\mathrm{i}k_{\mathrm{x}}\hat{p},
\end{equation}
\begin{equation}
\label{eq:lse_y_mom}
	\left(\sigma+\mathrm{i}k_{\mathrm{x}} U\right)\hat{v}+f\hat{u}=-\hat{p}_{y},
\end{equation}
\begin{equation}
\label{eq:lse_z_mom}
	\left(\sigma+\mathrm{i}k_{\mathrm{x}} U\right)\hat{w}=-\mathrm{i}k_{\mathrm{z}}\hat{p}+N^{2}\hat{T},
\end{equation}
\begin{equation}
\label{eq:lse_diffusion}
	\left(\sigma+\mathrm{i}k_{\mathrm{x}} U\right)\hat{T}+\hat{w}=\frac{1}{Pe}\hat{\nabla}^{2}\hat{T},
\end{equation}
where $\hat{\nabla}^{2}={\rm d}^{2}/{\rm d}y^{2}-k^{2}$ with $k^{2}=k_{\mathrm{x}}^{2}+k_{\mathrm{z}}^{2}$. Due to the symmetry $\sigma(k_{\mathrm{x}},k_{\mathrm{z}})=\sigma(k_{\mathrm{x}},-k_{\mathrm{z}})=\sigma(-k_{\mathrm{x}},k_{\mathrm{z}})^{*}=\sigma(-k_{\mathrm{x}},-k_{\mathrm{z}})^{*}$ where the asterisk $*$ denotes the complex conjugate; we thus consider only the positive wavenumbers $k_{\mathrm{x}}$ and $k_{\mathrm{z}}$ in this paper. 
For convenience and mathematical simplicity,
{we can simplify the set of equations (\ref{eq:lse_continuity}-\ref{eq:lse_diffusion}) into a single ordinary differential equation (ODE) for $\hat{T}$.
Firstly, from the vertical momentum and temperature equations, we express $\hat{w}$ and $\hat{p}$ as a function of $\hat{T}$:
\begin{equation}
\hat{w}=\left[-(\sigma+\mathrm{i}k_{\rm{x}}U)+\frac{1}{Pe}\hat{\nabla}^{2}\right]\hat{T},
\end{equation}
\begin{equation}
\label{eq:lse_p_T}
\hat{p}=\frac{\mathrm{i}}{k_{\rm{z}}}\left[-(\sigma+\mathrm{i}k_{\rm{x}}U)^{2}+\frac{\sigma+\mathrm{i}k_{\rm{x}}U}{Pe}\hat{\nabla}^{2}-N^{2}\right]\hat{T}.
\end{equation}
From the horizontal momentum equations (\ref{eq:lse_x_mom}-\ref{eq:lse_y_mom}), we can express $\hat{u}$ and $\hat{v}$ in terms of $\hat{p}$:
\begin{equation}
\hat{u}=\frac{(f-U')\frac{\mathrm{d}\hat{p}}{\mathrm{d}y}+\mathrm{i}k_{\rm{x}}(\sigma+\mathrm{i}k_{\rm{x}}U)\hat{p}}{f(U'-f)-(\sigma+\mathrm{i}k_{\rm{x}}U)^{2}},
\end{equation}
\begin{equation}
\hat{v}=\frac{(\sigma+\mathrm{i}k_{\rm{x}}U)\frac{\mathrm{d}\hat{p}}{\mathrm{d}y}-\mathrm{i}k_{\rm{x}}f\hat{p}}{f(U'-f)-(\sigma+\mathrm{i}k_{\rm{x}}U)^{2}}.
\end{equation}
Applying $\hat{p}$ of (\ref{eq:lse_p_T}) to $\hat{u}$ and $\hat{v}$, we have the velocity perturbation $\hat{{u}}=(\hat{u},\hat{v},\hat{w})$ expressed by $\hat{T}$ and the continuity equation (\ref{eq:lse_continuity}) becomes the single 4th-order ODE for $\hat{T}$:}
\begin{eqnarray}
\label{eq:lse_4thODE}
	&&\frac{{\rm d}^{2}\hat{T}}{{\rm d}y^{2}}+\left(\frac{4ss'}{s^{2}-N^{2}}-\frac{\Gamma'}{\Gamma}\right)\frac{{\rm d}\hat{T}}{{\rm d}y}\nonumber\\
	&&+\left[k_{\mathrm{z}}^{2}\frac{\Gamma}{N^{2}-s^{2}}-k_{\mathrm{x}}^{2}+f\frac{k_{\mathrm{x}}\Gamma'}{s\Gamma}+\frac{2ss'}{s^{2}-N^{2}}\left(\frac{U''}{U'}+\frac{s'}{s}-\frac{\Gamma'}{\Gamma}\right)\right]\hat{T}\nonumber\\
	&&+\frac{1}{Pe}\left(\frac{\mathrm{i}s}{s^{2}-N^{2}}\right)\left[\hat{\nabla}^{4}\hat{T}+\left(\frac{s'}{s}-\frac{\Gamma'}{\Gamma}\right)\hat{\nabla}^{2}\frac{{\rm d}\hat{T}}{{\rm d}y}+
	\right.\nonumber\\
	&&\left.\left\{\frac{s'}{s}\left(\frac{U''}{U'}-\frac{\Gamma'}{\Gamma}\right)+f\left(\frac{k_{\mathrm{x}}\Gamma'}{s\Gamma}-\frac{k_{\mathrm{z}}^{2}}{s^{2}}(U'-f)\right)\right\}\hat{\nabla}^{2}\hat{T}\right]=0,
\end{eqnarray}
where $\hat{\nabla}^{4}=(\hat{\nabla}^{2})^{2}$, $s=-\mathrm{i}\sigma+k_{\mathrm{x}} U$ is the complex Doppler-shifted frequency, and $\Gamma$ is the function defined as
\begin{equation}
\label{eq:lse_Gamma}
	\Gamma=s^{2}+f\left(U'-f\right).
\end{equation}
We note that Eq. (\ref{eq:lse_4thODE}) becomes the 2nd-order ODE in the nondiffusive limit $Pe\rightarrow\infty$ while it becomes independent of $N$ for the high-diffusivity case as $Pe\rightarrow0$. 
\subsection{Numerical method}
We solve the set of equations (\ref{eq:lse_continuity}-\ref{eq:lse_diffusion}) numerically by considering an eigenvalue problem in a simplified matrix form:
\begin{equation}
\label{eq:lse_matrix}
	\mathcal{A}
	\left(
	\begin{array}{c}
	\hat{u}\\
	\hat{v}\\
	\hat{T}
	\end{array}
	\right)=
	\sigma\mathcal{B}
	\left(
	\begin{array}{c}
	\hat{u}\\
	\hat{v}\\
	\hat{T}
	\end{array}
	\right),
\end{equation}
where $\mathcal{A}$ and $\mathcal{B}$ are operator matrices expressed as
\begin{equation}
\label{eq:matrixA}
	\mathcal{A}=
	\left[
	\begin{array}{ccc}
	\mathcal{A}_{11} & \mathcal{A}_{12} & 0\\
	\mathrm{i}k^{2}f & \mathcal{A}_{22} & N^{2}k_{\mathrm{z}}\frac{\mathrm{d}}{\mathrm{d}y}\\
	\mathrm{i}k_{\mathrm{x}} & \frac{\mathrm{d}}{\mathrm{d}y} & k_{\mathrm{x}} k_{\mathrm{z}} U+\frac{\mathrm{i}k_{\mathrm{z}}}{Pe}\hat{\nabla}^{2}
	\end{array}
	\right],
\end{equation}
\begin{equation}
\label{eq:matrixB}
	\mathcal{B}=
	\left[
	\begin{array}{ccc}
	-\frac{\mathrm{d}}{\mathrm{d}y} & \mathrm{i}k_{\mathrm{x}} & 0\\
	0 & \mathrm{i}\hat{\nabla}^{2} & 0\\
	0 & 0 & \mathrm{i}k_{\mathrm{z}}
	\end{array}
	\right],
\end{equation}
where
\begin{eqnarray}
\mathcal{A}_{11}&=&\mathrm{i}k_{\mathrm{x}}\left(U'+U\frac{\mathrm{d}}{\mathrm{d}y}\right)-\mathrm{i}k_{\mathrm{x}} f,\nonumber\\
\mathcal{A}_{12}&=&k_{\mathrm{x}}^{2}U+\left(U'-f\right)\frac{\mathrm{d}}{\mathrm{d}y}+U'',\nonumber\\
\mathcal{A}_{22}&=&k_{\mathrm{x}} U\hat{\nabla}^{2}+fk_{\mathrm{x}}\frac{\mathrm{d}}{\mathrm{d}y}-k_{\mathrm{x}} U''.
\end{eqnarray}
The eigenvalue problem (\ref{eq:lse_matrix}) is discretized in {the} $y$-direction using the rational Chebyshev functions with the mapping $\tilde{y}=y/\sqrt{1+y^{2}}$ projecting the Chebyshev domain $\tilde{y}\in(-1,1)$ onto the physical space $y\in (-\infty,\infty)$ \citep[]{Deloncle2007,Park2012}. 
Vanishing boundary conditions are imposed for $y\rightarrow\pm\infty$ by suppressing terms in the first and last rows of the operator matrices \citep[]{Antkowiak2005}. 
{We note that the eigenvalue problem (\ref{eq:lse_matrix}) is identical to that of \citet{Arobone2012} in the inviscid limit and when the density perturbation analogous to $-\hat{T}$ is considered.}
The numerical results are validated with those from \citet{Deloncle2007} and \citet{Arobone2012} in stratified and rotating fluids.

\section{General results}
%
%                                                One column figure
%----------------------------------------------------------------- 
   \begin{figure}
   \centering
   \includegraphics[width=6.5cm]{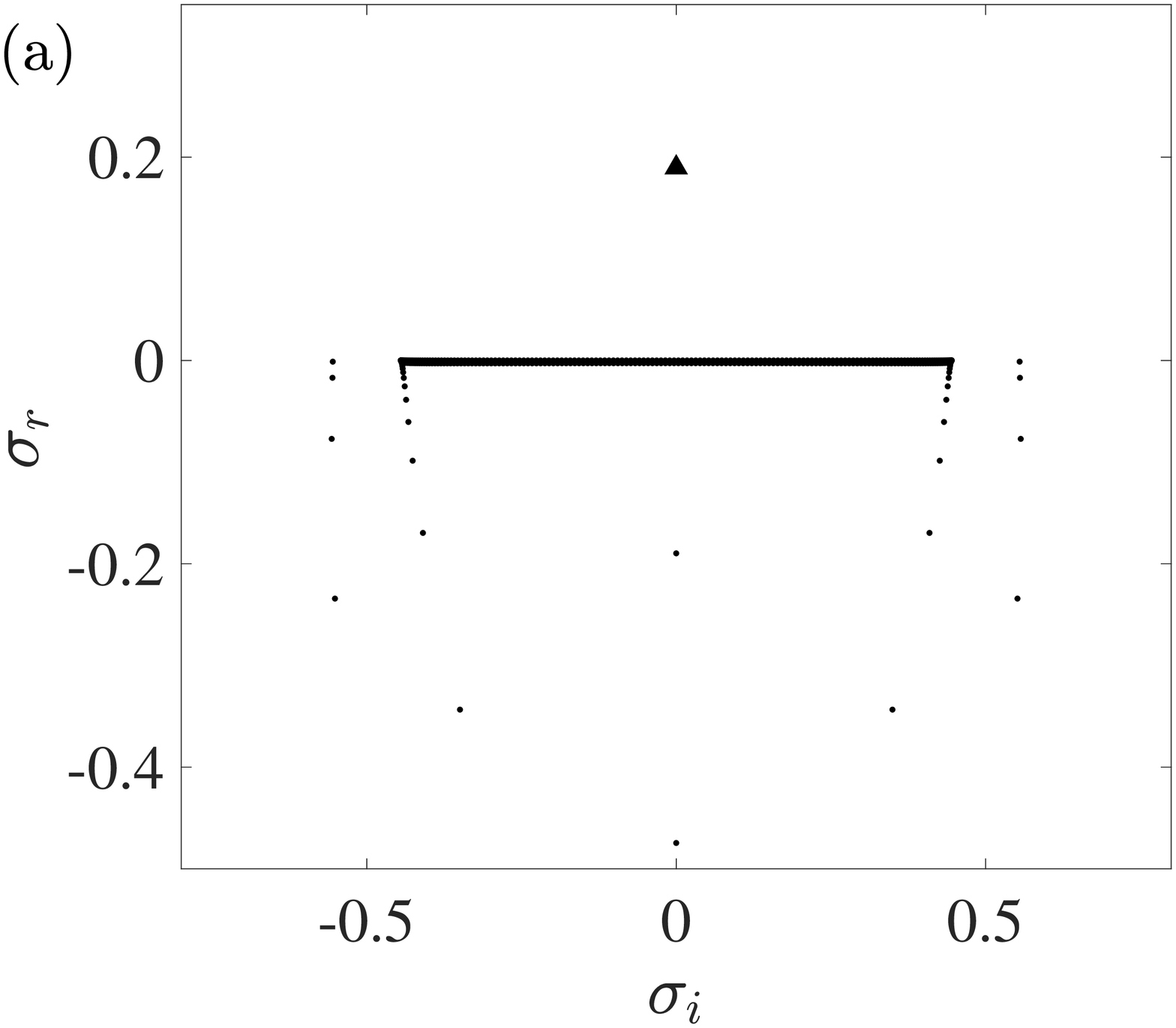}
      \includegraphics[width=6.5cm]{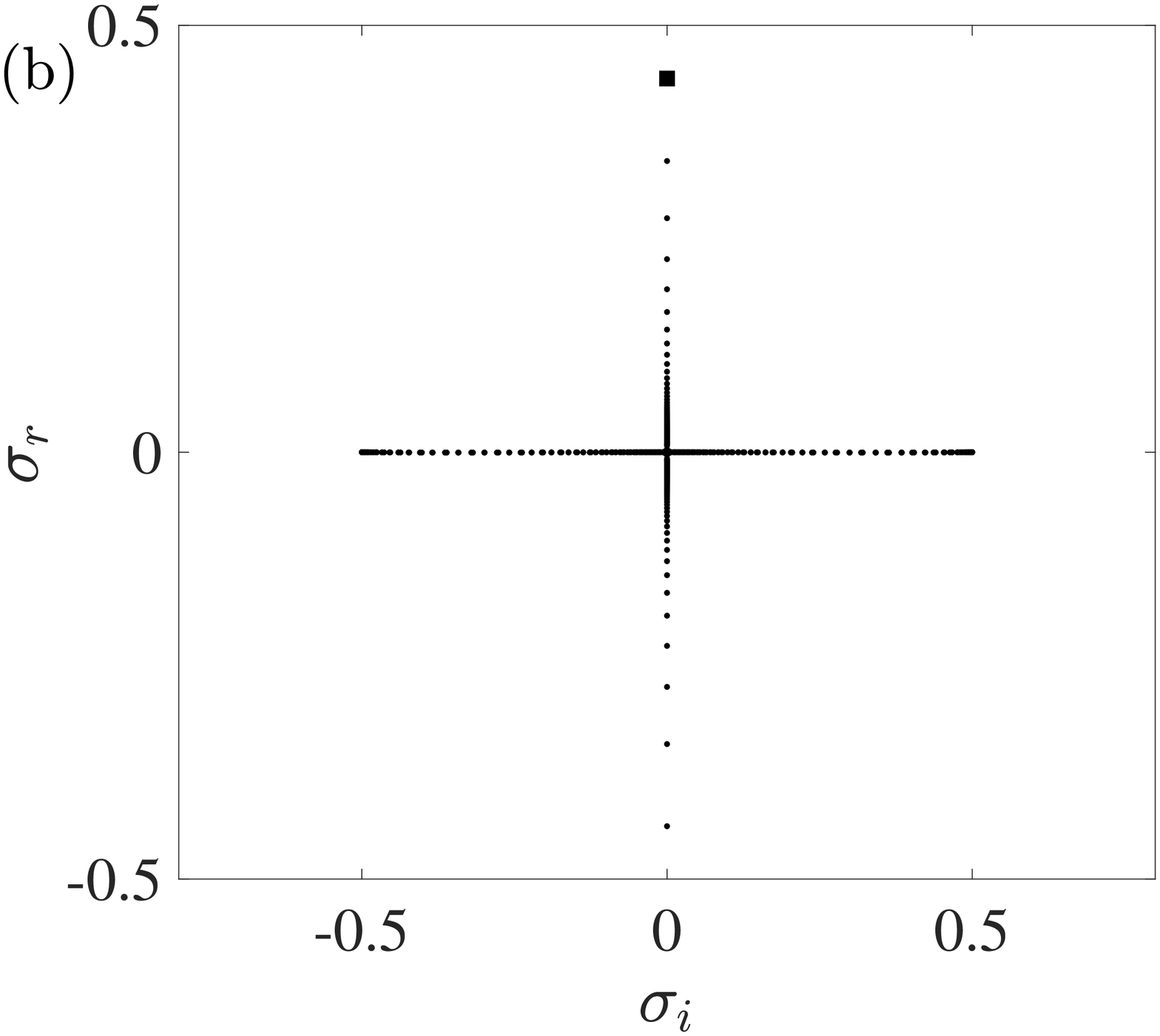}
     \caption{{Eigenvalue spectra at (a) $(k_{\mathrm{x}},k_{\mathrm{z}})=(0.445,0)$ and (b) $(k_{\mathrm{x}},k_{\mathrm{z}})=(0,5)$ for $f=0.5$, $N=1$, and $Pe=0.1$. The triangle in (a) and the square in (b) denote the most unstable growth rates of the inflectional and inertial instabilities, respectively. }
              }
         \label{Fig_eigenvalue_spectra}
   \end{figure}
%-------------------------------------- Two column figure (place early!)
   \begin{figure*}
   \centering
   \includegraphics[height=5.2cm]{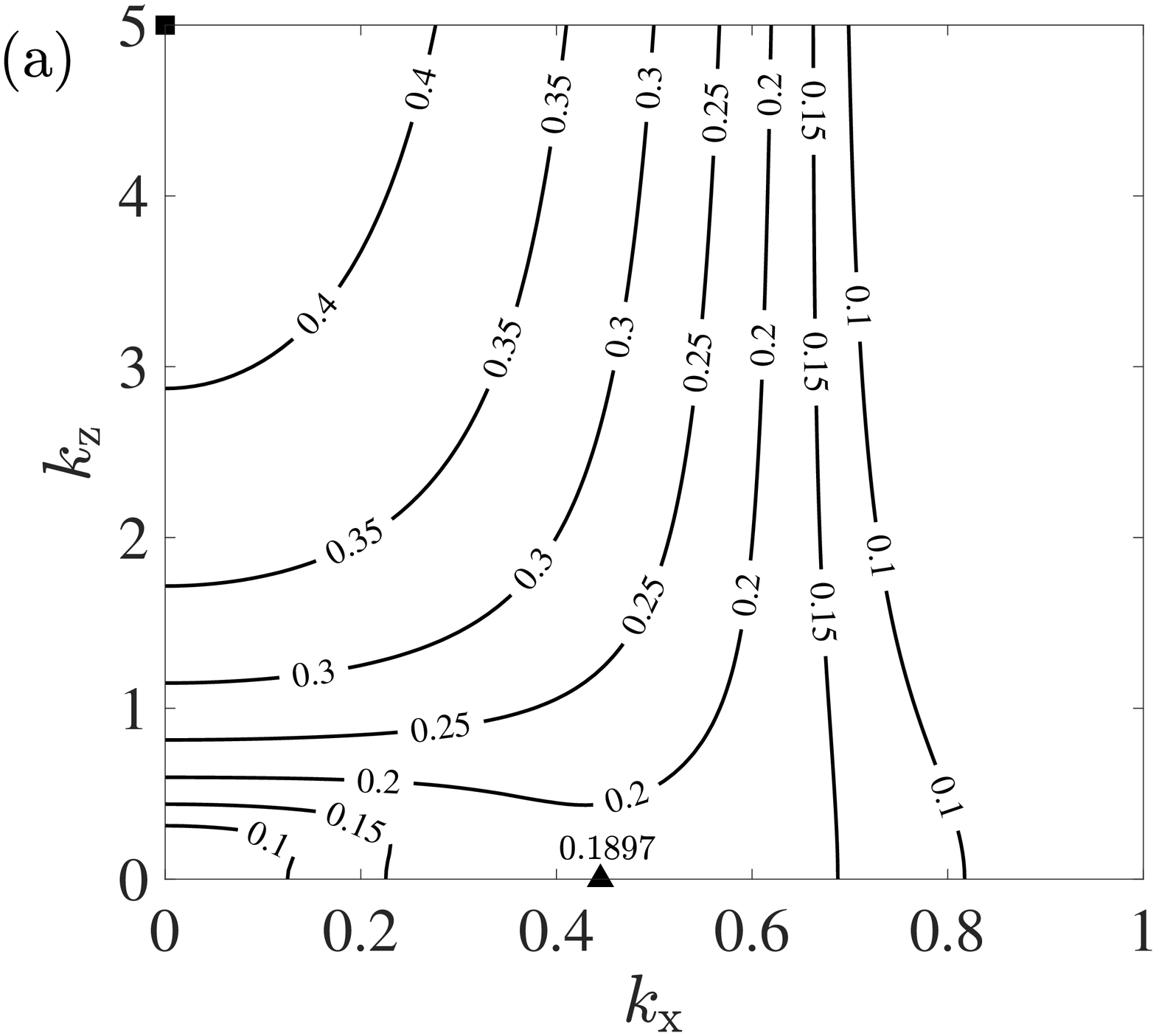}
   \includegraphics[height=5.2cm]{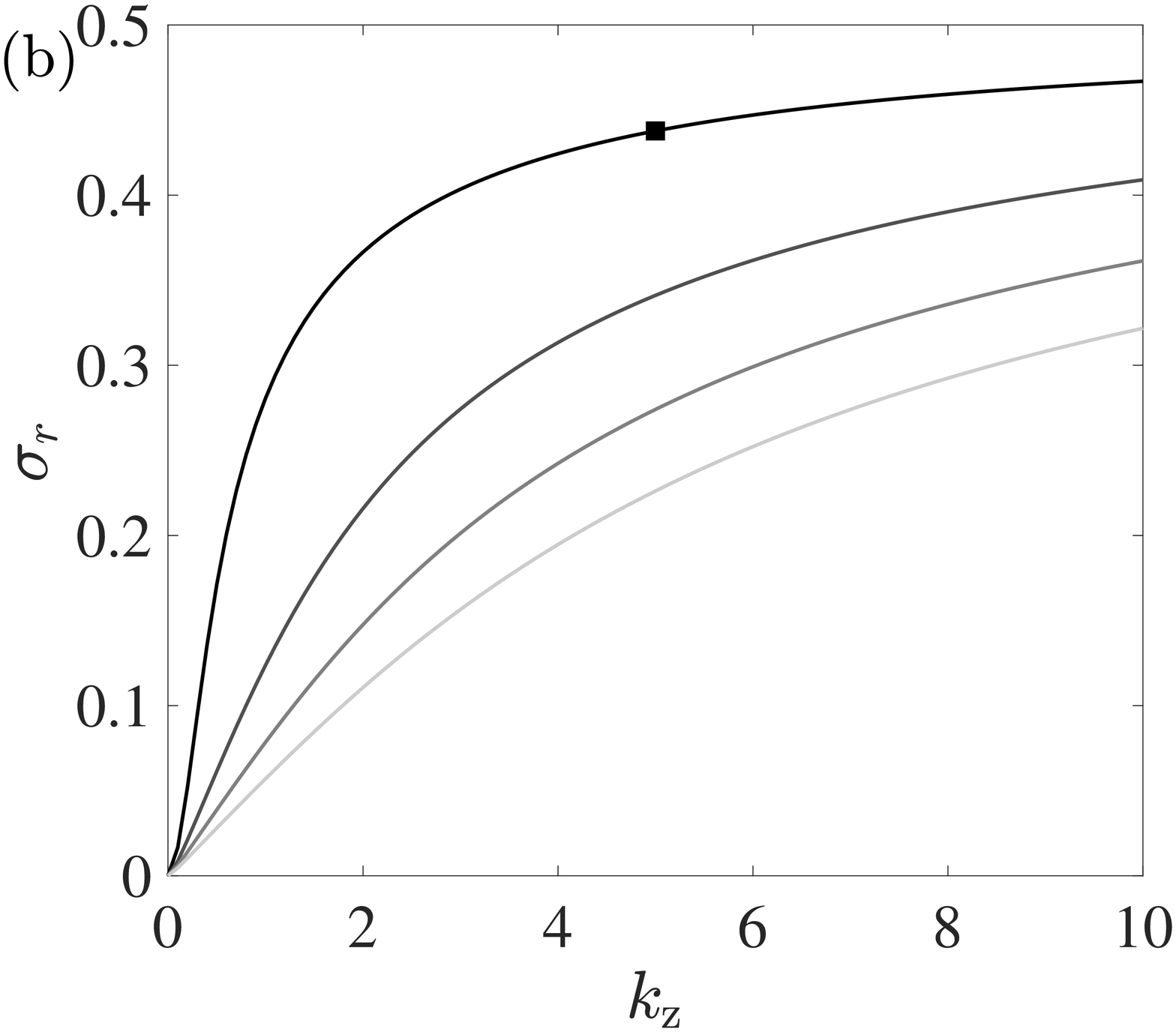}
   \includegraphics[height=5.2cm]{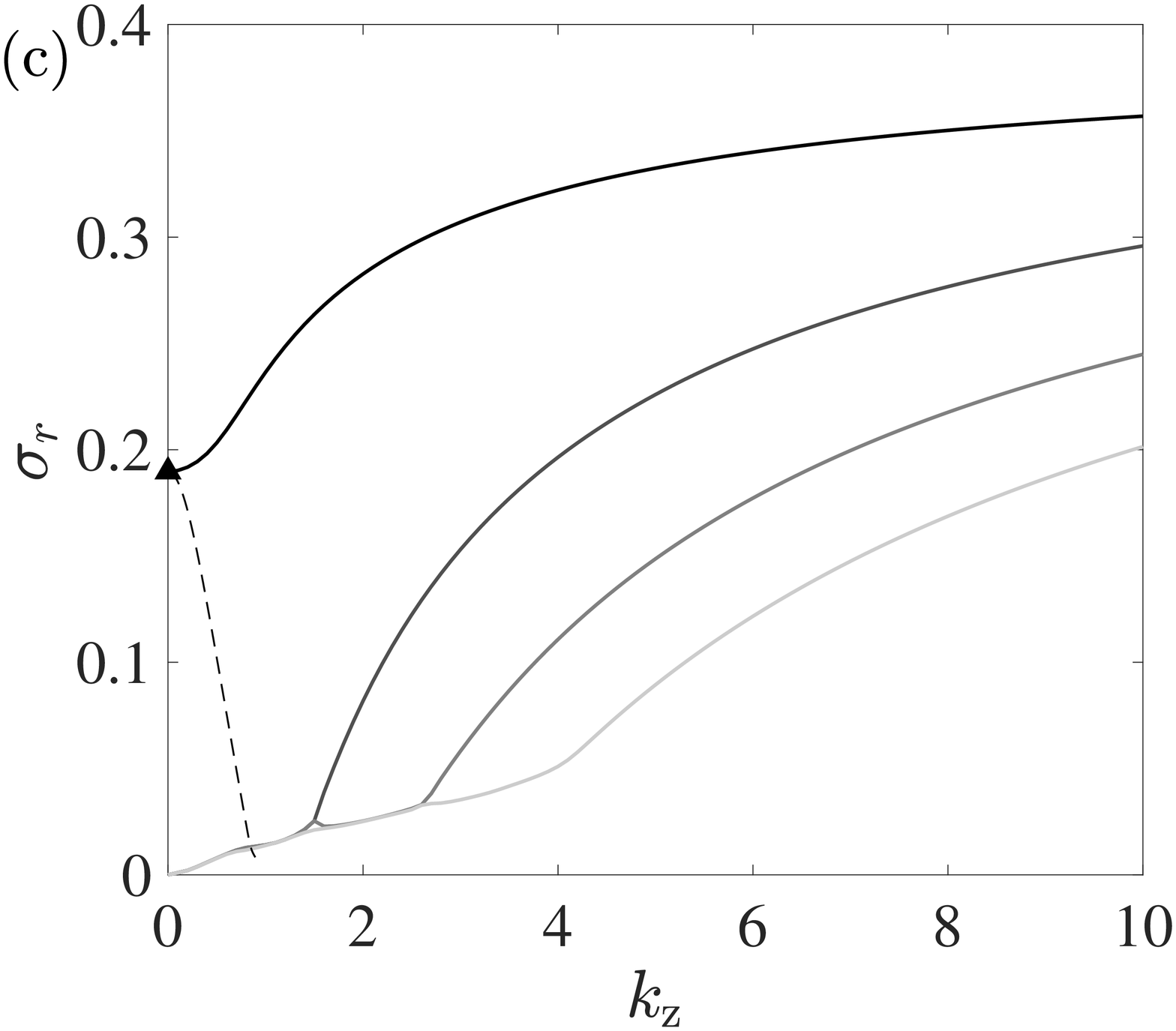}
   \caption{
   (a) Contours of the maximum growth rate $\sigma_{\rm max}$ in the parameter space of $(k_{\mathrm{x}},k_{\mathrm{z}})$ for $f=0.5$, $N=1$, $Pe=0.1$.   
   (b,c) Corresponding growth rates $\sigma_{r}$ versus vertical wavenumber $k_{\mathrm{z}}$ at (b) $k_{\mathrm{x}}=0$ and (c) $k_{\mathrm{x}}=0.445$. 
   For clarity, only the first four eigenvalue branches are plotted. The dashed line in (c) denotes the growth rate for rotating case $f=0$. 
   Symbols denote the parameters for {eigenvalues in Fig.~\ref{Fig_eigenvalue_spectra} and} eigenfunctions displayed in Fig.~\ref{Fig_general_2}.
   }
              \label{Fig_general_1}%
    \end{figure*}
 \begin{figure*}
   \centering
  \includegraphics[height=4cm]{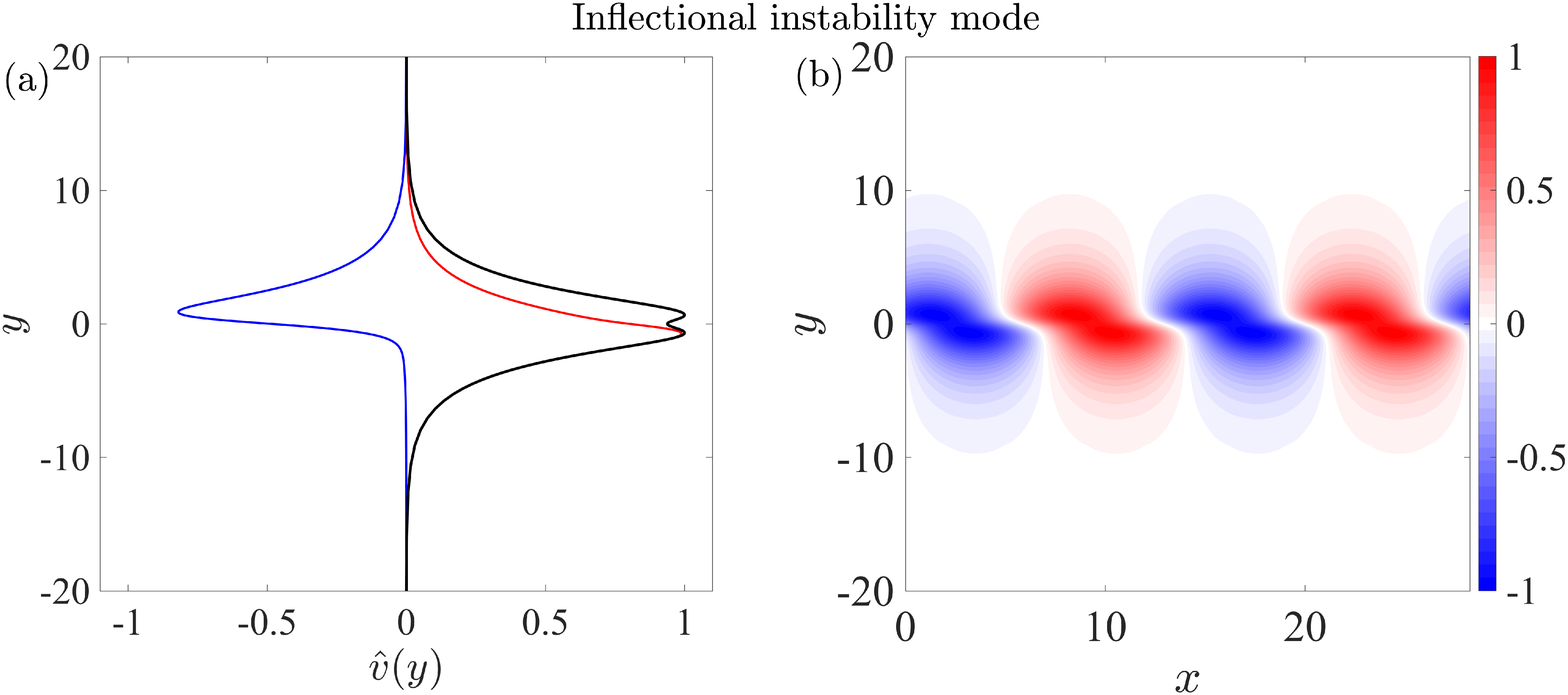}
  \includegraphics[height=4cm]{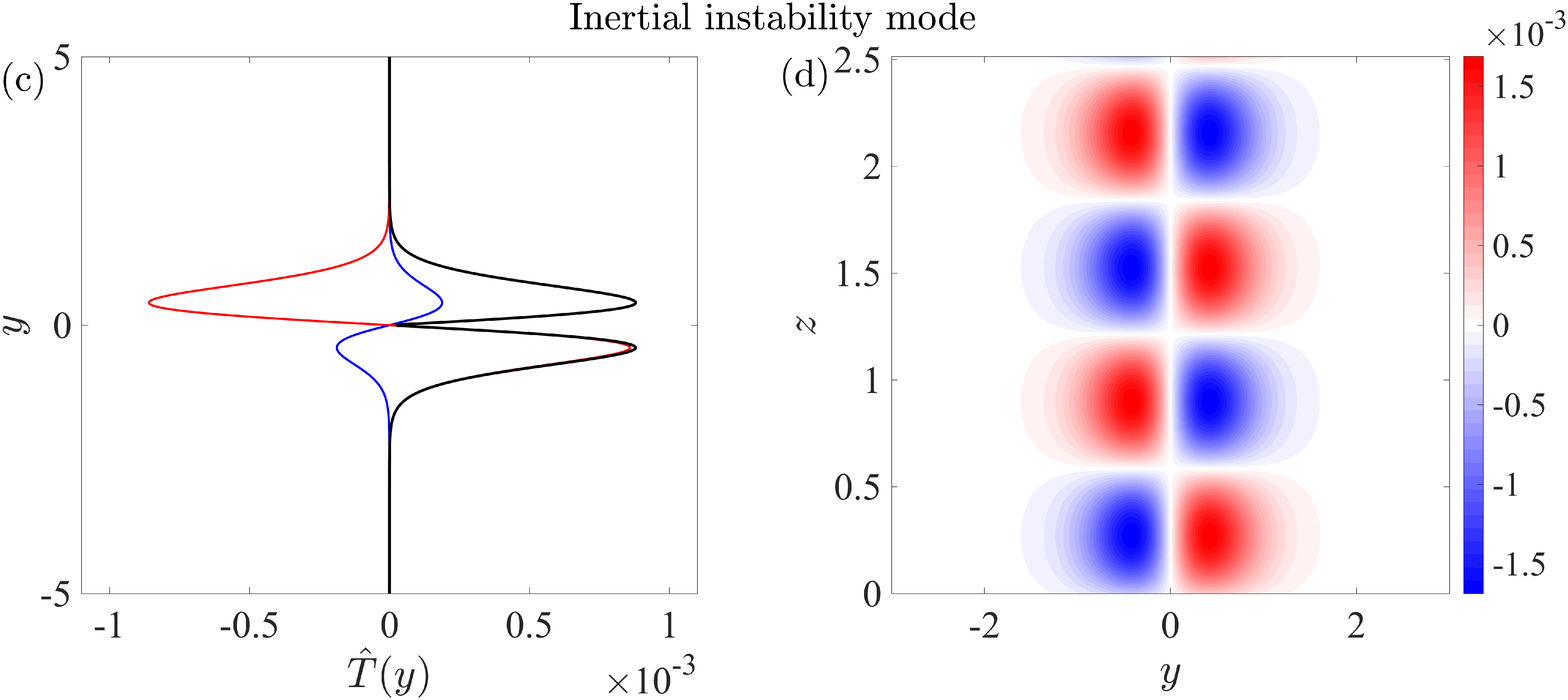}
    %%% Note for bounding box : [112 167 781 776]
     %%%\includegraphics{empty.eps}
   %%%\includegraphics{empty.eps}
   \caption{
   Examples of eigenmodes at (a,b) $(k_{\mathrm{x}},k_{\mathrm{z}})=(0.445,0)$ for the inflectional instability (triangle in Fig.~\ref{Fig_general_1}) and (c,d) $(k_{\mathrm{x}},k_{\mathrm{z}})=(0,5)$ for the inertial instability (square in Fig.~\ref{Fig_general_1}). (a) and (c) show the mode shapes of $\hat{v}(y)$ and $\hat{T}(y)$ (blue: real part, red: imaginary part, black: absolute value), respectively. (b) and (d) show the perturbations $\tilde{v}(x,y)$ and $\tilde{T}(y,z)$ in the physical space, respectively. 
      }
              \label{Fig_general_2}%
    \end{figure*}
Horizontal shear flows in stably stratified and rotating fluids are prone {to two types of destabilizing mechanisms}: the inflectional and inertial instabilities.
The inflectional instability occurs when there exists an inflection point $y_{i}$ where $U''(y_{i})=0$.
On the other hand, the inertial instability can occur without an inflection point when there is an imbalance between the pressure gradient and the inertial force, the mechanism equivalent for the centrifugal instability in cylindrical geometry \citep[][]{Kloosterziel1991,Billant2005,Wang2014}. 
For the hyperbolic tangent shear flow (\ref{eq:base_shear}) in stratified-rotating fluids, the inflectional instability is always present while the inertial instability exists only in the range $0<f<1$ \cite[]{Arobone2012}. 

Within this range, {we display in Fig.~\ref{Fig_eigenvalue_spectra} examples of eigenvalue spectra in the space $(\sigma_{i},\sigma_{r})$ at $f=0.5$, $N=1$, and $Pe=0.1$ for two sets of $(k_{\rm{x}},k_{\rm{z}})$: (a) $(0.445,0)$ and (b) $(0,5)$.
In Fig.~\ref{Fig_eigenvalue_spectra}a, the most unstable mode denoted by the triangle represents an inflectional instability mode. 
This mode has the growth rate $\sigma_{r}=0.1897$ and zero frequency $\sigma_{i}=0$ as equivalently reported in \cite{Deloncle2007} for the nondiffusive inflectional instability mode. 
At $k_{\mathrm{z}}=0$, we can derive a single 2nd-order ODE for $\hat{v}$ from the continuity equation (\ref{eq:lse_continuity}) and momentum equations (\ref{eq:lse_x_mom}-\ref{eq:lse_y_mom}):
\begin{equation}
\label{eq:2nd_ODE_v}
	\frac{{\rm d}^{2}\hat{v}}{{\rm d}y^{2}}-\left(k_{\mathrm{x}}^{2}+\frac{k_{\mathrm{x}} U''}{s}\right)\hat{v}=0,
\end{equation}
\citep[see also][]{Deloncle2007}.
It is important to note that the equation (\ref{eq:2nd_ODE_v}) does not depend on neither stratification, rotation, nor thermal diffusion, and the two-dimensional inflectional instability is consequently found to be independent of $N$  \citep[]{Deloncle2007} as well as $f$ and $Pe$.
We note that there also exist neutral waves (i.e., $\sigma_{r}=0$) with nonzero frequency lying collectively in the range $|\sigma_{i}|\lesssim k_{\rm{x}}=0.445$ as well as stable modes with $\sigma_{r}<0$.
In this study, however, we will consider only the unstable modes with zero frequency $\sigma_{i}=0$.
The eigenvalues in Fig.~\ref{Fig_eigenvalue_spectra}b at $(k_{\rm{x}},k_{\rm{z}})=(0,5)$ show different spectral behaviors from those in Fig.~\ref{Fig_eigenvalue_spectra}a: eigenvalues are distributed symmetrically with respect to $\sigma_{r}=0$ for unstable and stable modes while neutral waves lie in the frequency range $|\sigma_{i}|\lesssim f=0.5$. 
The most unstable mode denoted by the square corresponds to the inertial instability mode with the growth rate $\sigma_{r}=0.4378$ and zero frequency $\sigma_{i}=0$.
}

{In Fig.~\ref{Fig_general_1}a,} we display contours of the most unstable growth rate with thermal diffusion at $Pe=0.1$ to locate the inflectional and inertial instabilities in the parameter space $(k_{\mathrm{x}},k_{\mathrm{z}})$ for $f=0.5$ and $N=1$.
{The growth-rate contours are qualitatively similar to those in \citet{Arobone2012}, but we focus on the small-$Pe$ regime. 
The frequency} $\sigma_{i}$ of the most unstable mode is zero thus not plotted as contours in the parameter space $(k_{\mathrm{x}},k_{\mathrm{z}})$. 
Furthermore, we numerically verified that the inflectional instability is the most unstable in the two-dimensional case at $(k_{\mathrm{x}},k_{\mathrm{z}})=(0.445,0)$ for any values of $N$ and $Pe$ in the inertially-stable range ($f\geq1$ or $f\leq0$). On the other hand, in the inertially-unstable range $0<f<1$, the growth rate increases as $k_{\mathrm{z}}$ increases at $k_{\mathrm{x}}=0.445$ due to the inertial instability, and the most unstable growth rate of the inertial instability is found as $k_{\mathrm{z}}\rightarrow\infty$ at $k_{\mathrm{x}}=0$. 

Fig.~\ref{Fig_general_1}b shows an example of the growth rate $\sigma_{r}$ as a function of $k_{\mathrm{z}}$ at $k_{\mathrm{x}}=0$. 
{There is not only one unstable branch but a countless number of growth-rate branches (only the first four branches are shown in Fig.~\ref{Fig_general_1}b for clarity)}. 
We see that the first branch is the most unstable and all the branches asymptote {to} certain values as $k_{\mathrm{z}}$ increases. 
\citet{Arobone2012} argues that the inertial instability growth {rate approaches} $\sigma_{\rm max}=\sqrt{f(1-f)}$. 
In the next section, the reasons why there {is} {an infinite number} of branches and why they approach $\sqrt{f(1-f)}$ as $k_{\mathrm{z}}\rightarrow\infty$ will be explained by means of the WKBJ approximation. 
Fig.~\ref{Fig_general_1}c shows $\sigma_{r}$ versus $k_{\mathrm{z}}$ at $k_{\mathrm{x}}=0.445$. 
We see that the first branch starts from $\sigma_{r}=0.1897$ at $k_{\mathrm{z}}=0$ due to the inflectional instability, while other branches start at $\sigma_{r}=0$ and increase with $k_{\mathrm{z}}$. 
The increase of the growth rate $\sigma_{r}$ with $k_{\mathrm{z}}$ occurs only in the inertially-unstable regime ($0<f<1$) since the inflectional instability is stabilized as $k_{\mathrm{z}}$ increases in the inertially-stable regime (see {the dashed line} in Fig.~\ref{Fig_general_1}c for $f=0$).

{Fig.~\ref{Fig_general_2}} shows examples of modes for the inflectional instability at $(k_{\mathrm{x}},k_{\mathrm{z}})=(0.445,0)$ {in panels (a) and (b), and the inertial instability at $(k_{\mathrm{x}},k_{\mathrm{z}})=(0,5)$ in panels (c) and (d)} for the same parameters used in Fig.~\ref{Fig_general_1}. 
{For both instability modes}, the mode shapes are normalized by the maximum of $|\hat{v}|$. 
{For Fig.~\ref{Fig_general_2}, we show $\hat{v}$ as a representative for the inflectional instability to be able to compare with the previous literature \citep[]{Deloncle2007}.
For the inertial instability, we illustrate the behavior of $\hat{T}$ since this is the quantity we will use for its asymptotic analysis in Sect.~\ref{sec:WKBJ}.
The mode shape $\hat{v}$ of the inflectional instability decreases exponentially as $y\rightarrow\pm\infty$, and if plotted in physical space, we see that the perturbation $\tilde{v}(x,y)$ is slightly inclined against the direction of the shear \citep[see also][]{Arobone2012}}. 
On the other hand, the mode shape $\hat{T}$ of the inertial instability mode shows there exists a zero crossing for the absolute part of $\hat{T}$ at $y=0$ while decaying exponentially as $y\rightarrow\pm\infty$. 
The temperature perturbation $\tilde{T}$ in the space $(y,z)$ shows that the inertial instability mode has {a wave pattern} with a zero-crossing at $y=0$. One zero-crossing corresponds to the first branch, which is the most unstable branch for given $k_{\mathrm{x}}$ and $k_{\mathrm{z}}$, and higher branches have multiple zero-crossings accordingly.

\section{WKBJ analysis for the inertial instability}
\label{sec:WKBJ}
The maximum growth rate of the inertial instability is {proposed to be} $\sigma_{\rm max}=\sqrt{f(1-f)}$ \citep[]{Arobone2012}. 
We also verified numerically that the most unstable growth rate is reached as $k_{\mathrm{z}}\rightarrow\infty$ at $k_{\mathrm{x}}=0$. 
But it is still not clear why this maximum growth rate $\sigma_{\rm max}$ is attained as $k_{\mathrm{z}}\rightarrow\infty$, and when the inertial instability occurs in parameter ranges of $k_{\mathrm{z}}$, $N$, $f$, and $Pe$. 
In this section, we detail mathematical and physical interpretations of the inertial instability {using} the WKBJ approximation in the limit of large $k_{\mathrm{z}}$ at $k_{\mathrm{x}}=0$. Similar asymptotic analyses have been performed to understand the instability mechanisms in rotating shear flows \citep[]{Park2012JFM,Park2017}. 
To understand the effect of thermal diffusion, we perform an asymptotic analysis at two limits: $Pe\rightarrow\infty$ (i.e., no thermal diffusion) and $Pe\rightarrow0$ (i.e., high thermal diffusivity). 
It is noticeable that the diffusion is often neglected by taking $Pe\rightarrow\infty$ to understand geophysical flows in the atmosphere and oceans of the Earth \citep[]{Yavneh2001,Park2018}, while the high thermal diffusivity case with $Pe\rightarrow0$ is studied in astrophysical context for shear instability and mixing in stellar interiors \citep[]{Lignieresetal1999,PratLignieres2014}. 
In the following subsections, we provide explicit expressions of asymptotic dispersion relations for the inertial instability in stratified and rotating fluids in both limits of $Pe$.

\subsection{The weak diffusion limit: $Pe\rightarrow\infty$} 
We first consider the 4th-order ODE (\ref{eq:lse_4thODE}) for $k_{\mathrm{x}}=0$:
\begin{equation}
\begin{aligned}
\label{eq:lse_4thODE_alpha0}
	&\frac{\mathrm{d}^{4}\hat{T}}{\mathrm{d}y^{4}}-\frac{\Gamma'}{\Gamma}\frac{\mathrm{d}^{3}\hat{T}}{\mathrm{d}y^{3}}-\left[k_{\mathrm{z}}^{2}\left(1-\frac{\Gamma}{\sigma^{2}}\right)+Pe\left(\frac{N^{2}+\sigma^{2}}{\sigma}\right)\right]\frac{\mathrm{d}^{2}\hat{T}}{\mathrm{d}y^{2}}+\\
	&\left[k_{\mathrm{z}}^{2}\frac{\Gamma'}{\Gamma}+Pe\left(\frac{N^{2}+\sigma^{2}}{\sigma}\right)\frac{\Gamma'}{\Gamma}\right]\frac{\mathrm{d}\hat{T}}{\mathrm{d}y}-\left[k_{\mathrm{z}}^{4}\frac{\Gamma}{\sigma^{2}}+k_{\mathrm{z}}^{2}Pe\left(\frac{\Gamma}{\sigma}\right)\right]\hat{T}=0.
\end{aligned}
\end{equation}
{In the limit $Pe\rightarrow\infty$, we can rearrange equation (\ref{eq:lse_4thODE_alpha0}) as
\begin{equation}
\label{eq:lse_2ndODE_Pe_inf}
\frac{\mathrm{d}^{2}\hat{T}}{\mathrm{d}y^{2}}-\frac{\Gamma'}{\Gamma}\frac{\mathrm{d}\hat{T}}{\mathrm{d}y}+k_{\mathrm{z}}^{2}\frac{\Gamma}{N^{2}+\sigma^{2}}\hat{T}=O\left(\frac{1}{Pe}\right),
\end{equation}
where higher-order derivatives in the 3rd and 4th orders are also of the order $O\left(1/Pe\right)$. 
Provided that $\mathrm{d}/\mathrm{d}y$ is of order unity, we can neglect the term on the right-hand side as $Pe\rightarrow\infty$, and the equation (\ref{eq:lse_2ndODE_Pe_inf}) becomes the second-order ODE.}

Applying the WKBJ approximation to (\ref{eq:lse_2ndODE_Pe_inf}) for large $k_{\mathrm{z}}$:
\begin{equation}
\label{eq:WKBJ}
\hat{T}(y)\sim\exp\left[\frac{1}{\delta}\sum_{l=0}^{\infty}\delta^{l}S_{l}(y)\right],
\end{equation}
{we obtain}
\begin{equation}
\label{eq:WKBJ_S0S1}
\delta=\frac{1}{k_{\mathrm{z}}},~~~
S^{'2}_{0}=-\frac{\Gamma}{N^{2}+\sigma^{2}},~~~
S'_{1}=\frac{\Gamma'}{4\Gamma}.
\end{equation}
{In this paper, we only consider the inertial instability mode which has the zero frequency (i.e., $\sigma_{i}=0$ and $\sigma^{2}+N^{2}>0$) thus the sign of $\Gamma$ determines the behavior of the solutions.
We get evanescent solutions} if $\Gamma<0$:
\begin{equation}
\begin{aligned}
\label{eq:WKBJ_2ODE_exponential}
	\hat{T}(y)=(-\Gamma)^{1/4}&\left[A_{1}\exp\left(k_{\mathrm{z}}\int_{y}\sqrt{\frac{-\Gamma}{\sigma^{2}+N^{2}}}\mathrm{d}y\right)\right.\\
	&\left.+A_{2}\exp\left(-k_{\mathrm{z}}\int_{y}\sqrt{\frac{-\Gamma}{\sigma^{2}+N^{2}}}\mathrm{d}y\right)\right],
\end{aligned}
\end{equation}
where $A_{1}$ and $A_{2}$ are constants, or wavelike solutions if $\Gamma>0$:
\begin{equation}
\begin{aligned}
\label{eq:WKBJ_2ODE_wavelike}
	\hat{T}(y)=\Gamma^{1/4}&\left[B_{1}\exp\left(\mathrm{i}k_{\mathrm{z}}\int_{y}\sqrt{\frac{\Gamma}{\sigma^{2}+N^{2}}}\mathrm{d}y\right)\right.\\
	&\left.+B_{2}\exp\left(-\mathrm{i}k_{\mathrm{z}}\int_{y}\sqrt{\frac{\Gamma}{\sigma^{2}+N^{2}}}\mathrm{d}y\right)\right],
\end{aligned}
\end{equation}
where $B_{1}$ and $B_{2}$ are constants. 
These exponential behaviors depending on $y$ {change at} turning points $y_{t}$ where $\Gamma(y_{t})=0$. It is also convenient to introduce the turning growth rates $\sigma_{\pm}$:
\begin{equation}
\label{eq:def_turning_growthrate}
\sigma_{\pm}=\pm\Re\left[\sqrt{f(U'-f)}\right],
\end{equation}
which inform us where $\Gamma=-\sigma^{2}+f\left(U'-f\right)$ becomes zero.
%
%                                                One column figure
%----------------------------------------------------------------- 
   \begin{figure}
   \centering
   \includegraphics[height=5cm]{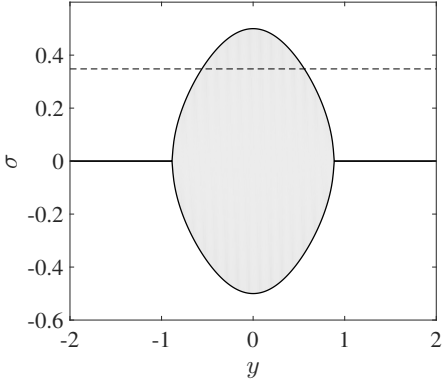}
      \caption{Turning growth rates $\sigma_{\pm}(y)$ for $f=0.5$ (black solid lines) and an example of the growth rate $\sigma=0.348$ at $(k_{\mathrm{x}},k_{\mathrm{z}})=(0,5)$ for $N=1$ and $Pe=\infty$ (dashed line). White and gray areas represent the regions where the solution is exponential and wavelike, respectively.  
              }
         \label{Fig_turning}
   \end{figure}
%-----------------------------------------------------------------
For instance, if the growth rate is $\sigma=0.348$ that is a growth rate obtained numerically for $f=0.5$, $N=1$, $Pe=\infty$, $k_{\mathrm{x}}=0$, and $k_{\mathrm{z}}=5$, there exist two turning points: one at $y=y_{t-}=-0.56$ and the other at $y=y_{t+}=0.56$ (see Fig.~\ref{Fig_turning}). 
This implies that the solution in the regions $y>y_{t+}$ and $y<y_{t-}$ is evanescent while the solution in the region $y_{t-}<y<y_{t+}$ is wavelike. 
We note that $\sigma_{\pm}=\pm\Re\left(\sqrt{-\Phi}\right)$ so we can verify that the region where the Rayleigh's discriminant $\Phi(y)=f(f-U')$ is negative lies between the two turning points $y_{t\pm}$.

When the growth rate lies between the two turning growth rates $\sigma_{-}<\sigma<\sigma_{+}$, this implies that the solutions are wavelike with $\Gamma>0$ (gray area in Fig.~\ref{Fig_turning}), while the solutions are evanescent outside this range (white area in Fig.~\ref{Fig_turning}). 
For other values of $\sigma$, we can decide whether we can construct eigenfunctions. 
For example, if $0<\sigma<\max(\sigma_{+})$ or $\min(\sigma_{-})<\sigma<0$ like the case $\sigma=0.4232<\max(\sigma_{+})=0.5$ in Fig.~\ref{Fig_general_2}(c,d), the solution is exponential outside the turning points $y_{t\pm}$ and wavelike in between. 
In this case, we can construct an eigenfunction which decays exponentially as $y\rightarrow\pm\infty$ due to the presence of two turning points. 
On the other hand, if the growth rate is either $\sigma>\max(\sigma_{+})$ or $\sigma<\min(\sigma_{-})$, then there is no turning point and solutions are always evanescent for all $y$. 
{Therefore, we must impose $A_{1}=0$ and $A_{2}=0$ to make the solution decaying with $y\rightarrow\infty$ and $y\rightarrow-\infty$, respectively, and no eigenfunction can be constructed in this case. }
{This also implies that the growth rate does not surpass $\max(\sigma_{+})=\sqrt{f(1-f)}$ to construct the inertial instability mode, which verifies the conjecture of \citet{Arobone2012}. }

{For an unstable mode that has the growth rate in the range $0<\sigma<\max(\sigma_{+})$,} we can further derive expressions for the asymptotic dispersion relation by performing a turning point analysis. We first consider the evanescent WKBJ solution that decays exponentially for $y>y_{t+}$:
\begin{equation}
\label{eq:WKBJ_solution_ytplus}
	\hat{T}(y)=A_{\infty}(-\Gamma)^{1/4}\exp\left(-k_{\mathrm{z}}\int_{y_{t+}}^{y}\sqrt{\frac{-\Gamma}{\sigma^{2}+N^{2}}}\mathrm{d}y\right),
\end{equation}
where $A_{\infty}$ is a constant. Around the turning point $y_{t+}$, the WKBJ solution (\ref{eq:WKBJ_solution_ytplus}) is no longer valid and we need to find a local solution that matches with the WKBJ solution in the range $y_{t-}<y<y_{t+}$. 
{By considering a new scaling $\tilde{y}=(y-y_{t+})/\epsilon$ and an approximation $\Gamma(y)\sim\Gamma'_{t+}\epsilon\tilde{y}$ where $\Gamma'_{t+}$ is the derivative of $\Gamma$ in $y$ at $y=y_{t+}$,} we obtain the following local equation
\begin{equation}
\label{eq:local_equation_ytplus}
\frac{\mathrm{d}^{2}\hat{T}}{\mathrm{d}\tilde{y}^{2}}-\frac{1}{\tilde{y}}\frac{\mathrm{d}\hat{T}}{\mathrm{d}\tilde{y}}-\tilde{y}\hat{T}=O(\epsilon),
\end{equation}
where $\epsilon=\left[k_{\mathrm{z}}^{2}(-\Gamma'_{t+})/(N^{2}+\sigma^{2})\right]^{-1/3}$. 
The solution of this local equation (\ref{eq:local_equation_ytplus}) can be expressed in terms of derivatives of the Airy functions: $\hat{T}(\tilde{y})=a_{1}\mathrm{Ai}'(\tilde{y})+b_{1}\mathrm{Bi}'(\tilde{y})$, where $a_{1}$ and $b_{1}$ are constants to be matched from the asymptotic behavior of the WKBJ solution (\ref{eq:WKBJ_solution_ytplus}) as $y\rightarrow y_{t+}$. 
From the asymptotic behaviors of the Airy functions when $\tilde{y}\rightarrow+\infty$:
\begin{equation}
\begin{aligned}
&\mathrm{Ai}'(\tilde{y})\rightarrow-\frac{\tilde{y}^{1/4}}{2\sqrt{\pi}}\exp\left(-\frac{2}{3}\tilde{y}^{3/2}\right),\\
&\mathrm{Bi}'(\tilde{y})\rightarrow\frac{\tilde{y}^{1/4}}{\sqrt{\pi}}\exp\left(\frac{2}{3}\tilde{y}^{3/2}\right),
\end{aligned}
\end{equation}
and when $\tilde{y}\rightarrow-\infty$:
\begin{equation}
\begin{aligned}
&\mathrm{Ai}'(\tilde{y})\rightarrow-\frac{(-\tilde{y})^{1/4}}{\sqrt{\pi}}\cos\left(\frac{2}{3}(-\tilde{y})^{3/2}+\frac{\pi}{4}\right),\\
&\mathrm{Bi}'(\tilde{y})\rightarrow\frac{(-\tilde{y})^{1/4}}{\sqrt{\pi}}\sin\left(\frac{2}{3}(-\tilde{y})^{3/2}+\frac{\pi}{4}\right),
\end{aligned}
\end{equation}
\citep[]{Abramowitz}, we obtain the matched WKBJ solution in the region $y_{t-}<y<y_{t+}$:
\begin{equation}
\begin{aligned}
\label{eq:WKBJ_2ODE_leftytplus}
	\hat{T}(y)=\Gamma^{1/4}&\left[C_{+}\exp\left(\mathrm{i}k_{\mathrm{z}}\int_{y_{t+}}^{y}\sqrt{\frac{\Gamma}{\sigma^{2}+N^{2}}}\mathrm{d}y\right)\right.\\
	&\left.+C_{-}\exp\left(-\mathrm{i}k_{\mathrm{z}}\int_{y_{t+}}^{y}\sqrt{\frac{\Gamma}{\sigma^{2}+N^{2}}}\mathrm{d}y\right)\right],
\end{aligned}
\end{equation}
where 
\begin{equation}
\label{eq:WKBJ_amplitudes}
	C_{+}=\exp\left(-\mathrm{i}\frac{\pi}{4}\right)A_{\infty},~~
	C_{-}=\exp\left(\mathrm{i}\frac{\pi}{4}\right)A_{\infty}.
\end{equation}
Similarly, {the evanescent solution in $y<y_{t-}$ that decays exponentially as $y\rightarrow-\infty$:}
\begin{equation}
\label{eq:WKBJ_solution_ytminus}
	\hat{T}(y)=A_{-\infty}(-\Gamma)^{1/4}\exp\left(k_{\mathrm{z}}\int_{y_{t-}}^{y}\sqrt{\frac{-\Gamma}{\sigma^{2}+N^{2}}}\mathrm{d}y\right),
\end{equation}
{matches with the following solution in $y_{t-}<y<y_{t+}$ after the local solution around $y_{t-}$ is considered:}
\begin{equation}
\begin{aligned}
\label{eq:WKBJ_2ODE_rightytminus}
	\hat{T}(y)=\Gamma^{1/4}&\left[B_{+}\exp\left(\mathrm{i}k_{\mathrm{z}}\int_{y_{t-}}^{y}\sqrt{\frac{\Gamma}{\sigma^{2}+N^{2}}}\mathrm{d}y\right)\right.\\
	&\left.+B_{-}\exp\left(-\mathrm{i}k_{\mathrm{z}}\int_{y_{t-}}^{y}\sqrt{\frac{\Gamma}{\sigma^{2}+N^{2}}}\mathrm{d}y\right)\right],
\end{aligned}
\end{equation}
where 
\begin{equation}
\label{eq:WKBJ_amplitudes_B}
	B_{+}=\exp\left(\mathrm{i}\frac{\pi}{4}\right)A_{-\infty},~~
	B_{-}=\exp\left(-\mathrm{i}\frac{\pi}{4}\right)A_{-\infty}.
\end{equation}
Matching the wavelike solutions (\ref{eq:WKBJ_2ODE_leftytplus}) and (\ref{eq:WKBJ_2ODE_rightytminus}):
\begin{equation}
\frac{B_{-}}{B_{+}}\exp\left(-2\mathrm{i}k_{\rm{z}}\int_{y_{t-}}^{y_{t+}}\sqrt{\frac{\Gamma}{\sigma^{2}+N^{2}}}\mathrm{d}y\right)=\frac{C_{-}}{C_{+}},
\end{equation}
that is
\begin{equation}
\exp\left(2\mathrm{i}k_{\rm{z}}\int_{y_{t-}}^{y_{t+}}\sqrt{\frac{\Gamma}{\sigma^{2}+N^{2}}}\mathrm{d}y\right)=\exp\left(-\mathrm{i}\pi\right),
\end{equation}
we obtain the dispersion relation in the form of a quantization formula:
\begin{equation}
\label{eq:WKBJ_dispersion_quantized}
	k_{\mathrm{z}}\int_{y_{t-}}^{y_{t+}}\sqrt{\frac{\Gamma}{\sigma^{2}+N^{2}}}\mathrm{d}y=\left(m-\frac{1}{2}\right)\pi,
\end{equation}
where $m$ is the branch number with a positive integer. This quantized dispersion relation implies that there exist {an infinite number} of discrete growth-rate branches for the inertial instability, as verified in {our} numerical results (Fig.~\ref{Fig_general_1}b,c). 

\begin{figure}
   \centering
   \includegraphics[height=6cm]{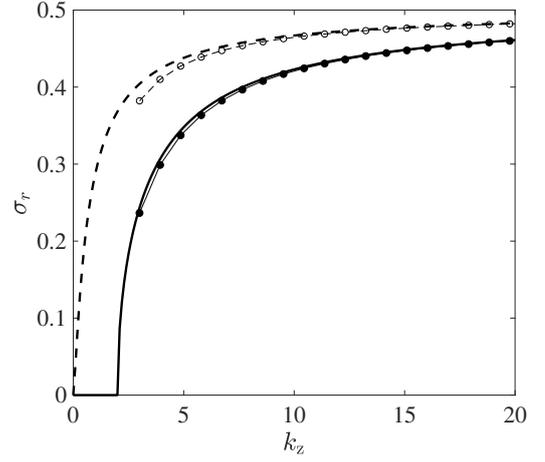}
      \caption{Growth rates of the first branch for $Pe=\infty$ (solid line) and $Pe=0.01$ (dashed line) at $f=0.5$ and $N=1$ with predictions from the WKBJ dispersion relations (\ref{eq:WKBJ_2ODE_dispersion_1st}) for $Pe=\infty$ (solid with filled circles) and (\ref{eq:WKBJ_4ODE_dispersion_1st}) for $Pe\rightarrow0$ (dashed with empty circles).
              }
         \label{Fig_growth_WKBJ}
   \end{figure}

The integral on the left-hand side of (\ref{eq:WKBJ_dispersion_quantized}) should become zero as $k_{\mathrm{z}}\rightarrow\infty$ since the right-hand side term is fixed with a finite $m$. 
This implies that the two turning points should approach each other as $k_{\mathrm{z}}$ goes to infinity while $\Gamma\rightarrow0$ (i.e., $y_{t-}$$\rightarrow$$y_{t+}$$\rightarrow0$). 
In addition to this property and the Taylor expansion of $\Gamma$ around $y=0$:
\begin{equation}
\Gamma\simeq-\sigma^{2}-f^{2}+fU'|_{y=0}+\frac{fU^{'''}|_{y=0}}{2}y^{2},
\end{equation}
we also assume that the growth rate can be expanded as $\sigma=\sigma_{0}-\sigma_{1}k_{\mathrm{z}}^{-p}+O(k_{\mathrm{z}}^{-2p})$ where $p$ is a positive integer.
Applying this growth-rate expansion to the dispersion relation (\ref{eq:WKBJ_dispersion_quantized}), we find that 
\begin{equation}
p=1,~~~
y^{2}_{t\pm}=\frac{2\sigma_{0}\sigma_{1}}{fk_{\rm{z}}},~~~
\sigma_{0}=\sqrt{f(1-f)},
\end{equation}
and we obtain the first-order term $\sigma_{1}$, which leads to a more explicit dispersion relation for very large $k_{\mathrm{z}}$:
\begin{equation}
\label{eq:WKBJ_2ODE_dispersion_1st}
	\sigma=\sigma_{0}-\frac{\sigma_{1}}{k_{\mathrm{z}}}+O\left(\frac{1}{k_{\mathrm{z}}^{2}}\right),
\end{equation}
where 
\begin{equation}
\label{eq:WKBJ_2ODE_dispersion_1st_each}
	\sigma_{1}=\frac{\sqrt{f}}{\sigma_{0}}\left(m-\frac{1}{2}\right)\sqrt{f(1-f)+N^{2}}.
\end{equation}
Since $\sigma_{1}$ is a positive value, we see that the maximum growth rate $\sigma_{\max}=\sigma_{0}=\sqrt{f(1-f)}$ is achieved at $k_{\mathrm{z}}\rightarrow\infty$ for any values of $m$. 
The dispersion relation (\ref{eq:WKBJ_2ODE_dispersion_1st}) also implies that the inertial instability only exists in the range $0<f<1$ to have a positive real $\sigma_{0}$. 
While the maximum growth rate $\sigma_{\rm max}$ is independent of $N$, the term $\sigma_{1}$ depends on the stratification and increases as $N$ increases; the growth rate thus decreases with $N$. 
We see in Fig.~\ref{Fig_growth_WKBJ} that the asymptotic dispersion relation (\ref{eq:WKBJ_2ODE_dispersion_1st}) matches well with the numerical results as $k_{\mathrm{z}}$ increases.

%
%                                                One column figure
%----------------------------------------------------------------- 
%-----------------------------------------------------------------
%
%                                                One column figure
%----------------------------------------------------------------- 
   \begin{figure}
   \centering
   \includegraphics[height=6cm]{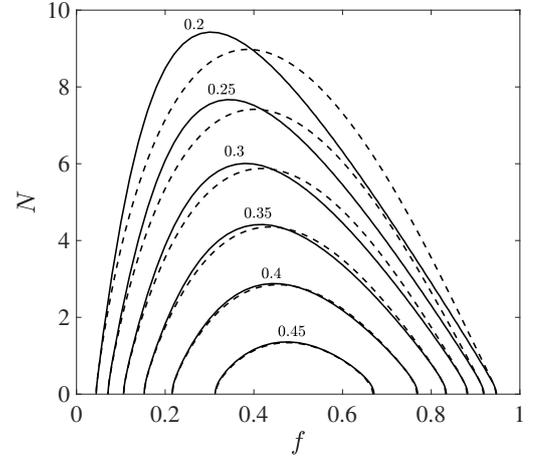}
      \caption{Growth rate contours in the parameter space $(f,N)$ for $Pe=\infty$, $k_{\mathrm{x}}=0$, $k_{\mathrm{z}}=20$ (solid lines), and predictions from the WKBJ dispersion relation (\ref{eq:WKBJ_2ODE_dispersion_1st}) (dashed lines).
              }
         \label{Fig_growth_WKBJ_contours}
   \end{figure}
%-----------------------------------------------------------------
The asymptotic dispersion relation (\ref{eq:WKBJ_2ODE_dispersion_1st}) {is a rich source of information regarding the inertial instability that covers a wide range of parameters $f$ and $N$.} 
In Fig.~\ref{Fig_growth_WKBJ_contours}, we see how the growth rate contours change in the parameter space $(f,N)$ for $Pe=\infty$, $k_{\mathrm{x}}=0$, and $k_{\mathrm{z}}=20$. 
At a fixed $f$, the growth rate decreases with $N$. 
The contours of the asymptotic growth rate match very well with {our} numerical results especially for higher values of the growth rate. 
{The better agreement at higher growth rates follows from our expansion of (\ref{eq:WKBJ_dispersion_quantized}).
Specifically, the turning points are assumed to be close to each other, which is equivalent to assuming that the growth rate is close to its maximum value.}

\subsection{The strong diffusion limit: $Pe\rightarrow 0$}
Now we investigate the limit $Pe\rightarrow0$ for high thermal diffusivity. The 4th-order ODE (\ref{eq:lse_4thODE_alpha0}) can be expressed as
\begin{equation}
\label{eq:lse_4thODE_Pe0}
	\frac{\mathrm{d}^{4}\hat{T}}{\mathrm{d}y^{4}}-\frac{\Gamma'}{\Gamma}\frac{\mathrm{d}^{3}\hat{T}}{\mathrm{d}y^{3}}-k_{\mathrm{z}}^{2}\left(1-\frac{\Gamma}{\sigma^{2}}\right)\frac{\mathrm{d}^{2}\hat{T}}{\mathrm{d}y^{2}}+k_{\mathrm{z}}^{2}\frac{\Gamma'}{\Gamma}\frac{\mathrm{d}\hat{T}}{\mathrm{d}y}-k_{\mathrm{z}}^{4}\frac{\Gamma}{\sigma^{2}}\hat{T}=O(Pe).
\end{equation}
Applying the WKBJ approximation (\ref{eq:WKBJ}), we obtain {the following relations at leading order:}
\begin{equation}
\delta=\frac{1}{k_{\mathrm{z}}},~~
S_{0}^{'4}-\left(1-\frac{\Gamma}{\sigma^{2}}\right)S_{0}^{'2}-\frac{\Gamma}{\sigma^{2}}=0,
\end{equation}
and at first order:
\begin{equation}
S'_{1}\left[4S_{0}^{'2}-2\left(1-\frac{\Gamma}{\sigma^{2}}\right)\right]+\frac{S_{0}^{''}}{S'_{0}}\left(6S_{0}^{'2}-1+\frac{\Gamma}{\sigma^{2}}\right)+\frac{\Gamma'}{\Gamma}(1-S_{0}^{'2})=0.
\end{equation}
The terms $S_{0}$ and $S_{1}$ satisfy
\begin{equation}
S_{0}^{'2}=-\frac{\Gamma}{\sigma^{2}},~~~
S_{1}^{'}=\frac{\Gamma'}{4\Gamma}-\frac{\Gamma'}{\Gamma+\sigma^{2}},
\end{equation}
{or}
\begin{equation}
S_{0}^{'2}=1,~~~
S_{1}^{'}=0.
\end{equation}
The general WKBJ solution with $S_{0}^{'2}(y)=1$ is 
\begin{equation}
\hat{T}(y)=D_{3}\exp(k_{\mathrm{z}} y)+D_{4}\exp(-k_{\mathrm{z}} y),
\end{equation}
but it has no turning point and the two solutions are always exponentially decaying or increasing as $y\rightarrow\pm\infty$. Therefore, we must impose $D_{3}=D_{4}=0$ to construct an eigenfunction. With $S_{0}^{'2}=-\Gamma/\sigma^{2}$, the WKBJ solution is wavelike if $\Gamma>0$:
\begin{equation}
\label{eq:WKBJ_4thODE_wave}
\begin{aligned}
\hat{T}(y)=\frac{\Gamma^{1/4}}{\left|\sigma^{2}+\Gamma\right|}&\left[D_{1}\exp\left(\mathrm{i}k_{\mathrm{z}}\int_{y}\frac{\sqrt{\Gamma}}{\sigma}\mathrm{d}y\right)\right.\\
&\left.+D_{2}\exp\left(-\mathrm{i}k_{\mathrm{z}}\int_{y}\frac{\sqrt{\Gamma}}{\sigma}\mathrm{d}y\right)\right],
\end{aligned}
\end{equation}
or evanescent if $\Gamma<0$:
\begin{equation}
\begin{aligned}
\label{eq:WKBJ_4thODE_exp}
\hat{T}(y)=\frac{(-\Gamma)^{1/4}}{\left|\sigma^{2}+\Gamma\right|}&\left[E_{1}\exp\left(k_{\mathrm{z}}\int_{y}\frac{\sqrt{-\Gamma}}{\sigma}\mathrm{d}y\right)\right.\\
&\left.+E_{2}\exp\left(-k_{\mathrm{z}}\int_{y}\frac{\sqrt{-\Gamma}}{\sigma}\mathrm{d}y\right)\right].
\end{aligned}
\end{equation}
These solutions in (\ref{eq:WKBJ_4thODE_wave}) and (\ref{eq:WKBJ_4thODE_exp}) now have turning points where $\Gamma=0$, which is the same as in the case without thermal diffusion at $Pe=\infty$. Therefore, the turning growth rates $\sigma_{\pm}(y)$ of the solutions (\ref{eq:WKBJ_4thODE_wave}-\ref{eq:WKBJ_4thODE_exp}) are essentially the same as the expression (\ref{eq:def_turning_growthrate}), and we can construct an eigenfunction if the growth rate $\sigma$ lies in the ranges $\min(\sigma_{-})<\sigma<0$ or $0<\sigma<\max(\sigma_{+})$. 

For $Pe\rightarrow0$, we can also perform a turning point analysis in order to obtain the asymptotic dispersion relation. We consider the WKBJ solution that decays exponentially for $y>y_{t+}$:
\begin{equation}
\hat{T}(y)=E_{\infty}\frac{(-\Gamma)^{1/4}}{\left|\sigma^{2}+\Gamma\right|}\exp\left(-k_{\mathrm{z}}\int_{y_{t+}}^{y}\frac{\sqrt{-\Gamma}}{\sigma}\mathrm{d}y\right).
\end{equation}
Around the turning point $y_{t+}$, we apply a new scaled coordinate $\tilde{y}=(y-y_{t+})/\epsilon$ and obtain the following local equation
\begin{equation}
\epsilon\frac{\mathrm{d}^{4}\hat{T}}{\mathrm{d}\tilde{y}^{4}}-\frac{\epsilon}{\tilde{y}}\frac{\mathrm{d}^{3}\hat{T}}{\mathrm{d}\tilde{y}^{3}}-k_{\mathrm{z}}^{2}\epsilon^{3}\frac{\mathrm{d}^{2}\hat{T}}{\mathrm{d}\tilde{y}^{2}}+\frac{k_{\mathrm{z}}^{2}\epsilon^{3}}{\tilde{y}}\frac{\mathrm{d}\hat{T}}{\mathrm{d}\tilde{y}}+\frac{k_{\mathrm{z}}^{4}\epsilon^{6}(-\Gamma'_{t+})}{\sigma^{2}}\hat{T}=0.
\end{equation}
Here, we take $\epsilon=\left[k_{\mathrm{z}}^{2}(-\Gamma'_{t+})/\sigma^{2}\right]^{-1/3}$ to balance the equation, and the 3rd- and 4th-order derivatives become negligible since they are of order $O(\epsilon)$. 
The final local equation becomes the same as (\ref{eq:local_equation_ytplus}) thus the local solution is the sum of derivatives of the Airy functions. 
By matching the solution around $y_{t+}$ {using the asymptotic behaviors of the Airy functions as $\tilde{y}\rightarrow\pm\infty$ \citep[]{Abramowitz} similar to in the previous section}, we obtain the WKBJ solution for $y_{t-}<y<y_{t+}$
\begin{equation}
\label{eq:WKBJ_4ODE_leftytplus}
\begin{aligned}
\hat{T}(y)=\frac{\Gamma^{1/4}}{\left|\sigma^{2}+\Gamma\right|}&\left[F_{+}\exp\left(\mathrm{i}k_{\mathrm{z}}\int_{y_{t+}}^{y}\frac{\sqrt{\Gamma}}{\sigma}\mathrm{d}y\right)\right.\\
&\left.+F_{-}\exp\left(-\mathrm{i}k_{\mathrm{z}}\int_{y_{t+}}^{y}\frac{\sqrt{\Gamma}}{\sigma}\mathrm{d}y\right)\right],
\end{aligned}
\end{equation}
where 
\begin{equation}
\label{eq:WKBJ_4ODE_amplitudes}
	F_{+}=\exp\left(-\mathrm{i}\frac{\pi}{4}\right)E_{\infty},~~
	F_{-}=\exp\left(\mathrm{i}\frac{\pi}{4}\right)E_{\infty}.
\end{equation}
By matching the solution (\ref{eq:WKBJ_4ODE_leftytplus}) with the solution for $y<y_{t-}$:
\begin{equation}
\hat{T}(y)=E_{-\infty}\frac{(-\Gamma)^{1/4}}{\left|\sigma^{2}+\Gamma\right|}\exp\left(k_{\mathrm{z}}\int_{y_{t-}}^{y}\frac{\sqrt{-\Gamma}}{\sigma}\mathrm{d}y\right),
\end{equation}
we obtain the following dispersion relation in a quantization form
\begin{equation}
\label{eq:WKBJ_4ODE_dispersion_quantized}
	k_{\mathrm{z}}\int_{y_{t-}}^{y_{t+}}\frac{\sqrt{\Gamma}}{\sigma}\mathrm{d}y=\left(m_{0}-\frac{1}{2}\right)\pi,
\end{equation}
where $m_{0}$ is the positive integer indicating the branch number. Similarly to the previous section, we Taylor-expand the growth rate $\sigma$ and we obtain an explicit asymptotic dispersion relation for $\sigma$ as a function of $k_{\mathrm{z}}$:
\begin{equation}
\label{eq:WKBJ_4ODE_dispersion_1st}
\sigma=\sigma_{0,0}-\frac{\sigma_{1,0}}{k_{\mathrm{z}}}+O\left(\frac{1}{k_{\mathrm{z}}^{2}}\right),
\end{equation}
where
\begin{equation}
\label{eq:WKBJ_4ODE_dispersion_1st_each}
\sigma_{0,0}=\sqrt{f(1-f)},~~\sigma_{1,0}=\left(m_{0}-\frac{1}{2}\right)\sqrt{f}.
\end{equation}
We see in Fig.~\ref{Fig_growth_WKBJ} that a numerical result at small $Pe=0.01$ matches very well with the asymptotic growth rate (\ref{eq:WKBJ_4ODE_dispersion_1st}) in the limit $Pe\rightarrow0$ as $k_{\mathrm{z}}$ increases. It is important to note that the expression of the maximum growth rate is the same as (\ref{eq:WKBJ_2ODE_dispersion_1st}) in nondiffusive fluids, and the inertial instability in stratified-rotating fluids with high thermal diffusivity also occurs in the regime $0<f<1$. In contrast to the nondiffusive case, the term $\sigma_{1,0}$ is independent of the stratification $N$.

\subsection{Comparison of the growth rates}
It is noteworthy that the asymptotic dispersion relation (\ref{eq:WKBJ_4ODE_dispersion_1st}) in fluids with high thermal diffusivity is independent of the stratification, and the expression (\ref{eq:WKBJ_2ODE_dispersion_1st}) without thermal diffusion becomes identical to (\ref{eq:WKBJ_4ODE_dispersion_1st}) as $N\rightarrow0$. This implies that high thermal diffusivity suppresses the effect of the stable stratification. Thus, stratified fluids with high thermal diffusivity behave like unstratified fluids. The term $\sigma_{1,0}$ at the first order is always smaller than $\sigma_{1}$ in (\ref{eq:WKBJ_2ODE_dispersion_1st_each}); therefore, the ratio $\gamma_{1}$ between the two terms at first order is always larger than unity for positive $N$ if we consider the same branch $m=m_{0}$:
\begin{equation}
\label{eq:ratio_1st_order}
	\gamma_{1}=\frac{\sigma_{1}}{\sigma_{1,0}}=\sqrt{\frac{f(1-f)+N^{2}}{f(1-f)}}>1.
\end{equation}
The growth rate of inertial instability in thermally-diffusive fluids is therefore always larger than that in nondiffusive fluids. The effect of high thermal diffusivity has already been reported in previous literature for the vertical shear instability \citep[]{Lignieres1999,PratLignieres2013} but the above expression shows explicitly and quantitatively how much the inertial instability is destabilized by {a high thermal diffusivity} as shown in Fig.~\ref{Fig_growth_WKBJ}.
\section{Parametric study}
\subsection{Effects of $Pe$ on the inflectional instability}
\begin{figure*}
   \centering
   \includegraphics[height=5cm]{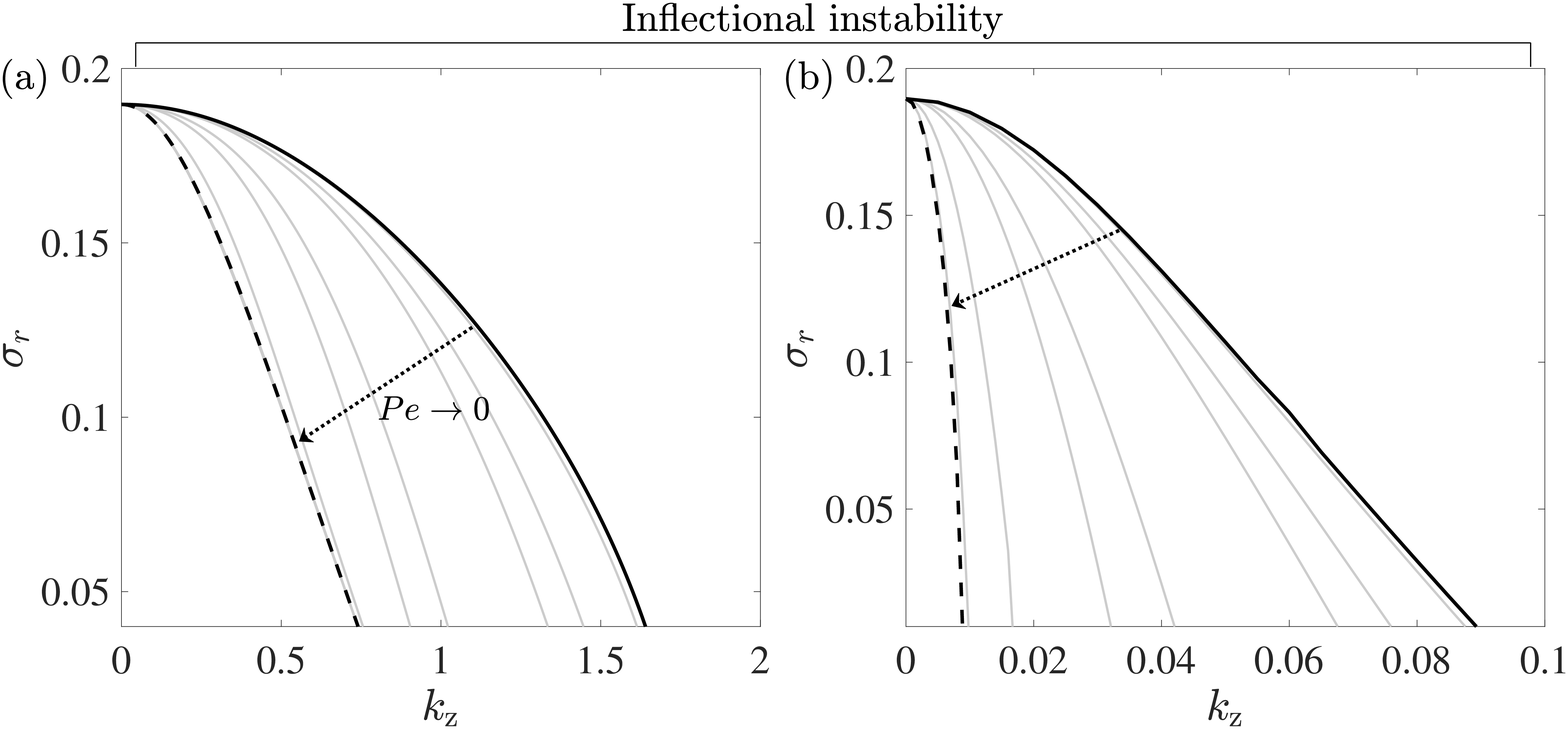}
   \includegraphics[height=5cm]{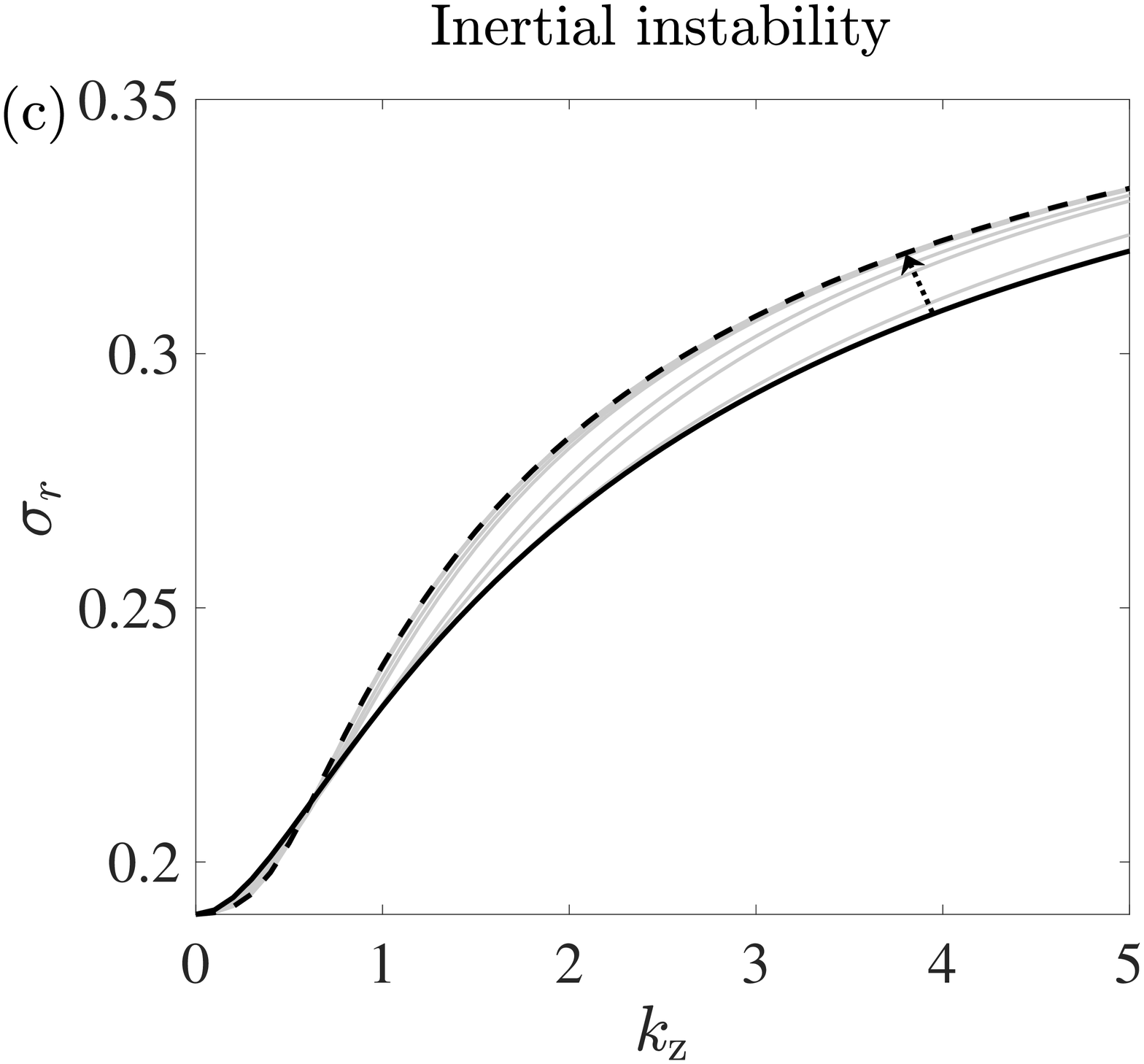}
   \caption{
   Growth rate $\sigma_{r}$ versus vertical wavenumber $k_{\mathrm{z}}$ at $k_{\mathrm{x}}=0.445$ for different P\'eclet numbers $Pe$: $Pe=\infty$ (black solid line), $Pe=\left\{100,10,5,1,0.5,0.1,0.01\right\}$ (gray solid lines) and other parameters: (a) $N=2$, $N/f=\infty$, (b) $N=2$, $N/f=0.1$, (c) $N=0.5$, $N/f=1$. Dashed lines denote the growth rate for unstratified case $N=0$ {at (a) $f=0$, (b) $f=20$, (c) $f=0.5$}. Arrows indicate the direction of decreasing $Pe$. 
   }
              \label{Fig_growth_Pe}%
    \end{figure*}
While the inertial instability is accessible by the WKBJ approximation, the inflectional instability should be investigated numerically since it occurs at low $k_{\mathrm{x}}$ and $k_{\mathrm{z}}$. 
In Fig.~\ref{Fig_growth_Pe}a, we see the growth rate $\sigma_{r}$ of the inflectional instability as a function of $k_{\mathrm{z}}$ for different values of the P\'eclet number $Pe$ at $k_{\mathrm{x}}=0.445$, $N=2$, and $f=0$, the regime where only the inflectional instability exists. 
{While the two-dimensional growth rate at $k_{\rm{z}}=0$ remains the same regardless of $Pe$, the} growth rate $\sigma_{r}$ decreases with $k_{\mathrm{z}}$ for any values of $Pe$, and the stabilization occurs faster as $Pe$ decreases to zero (i.e., as the thermal diffusivity increases). 
We see that growth-rate curves (solid lines) approach the growth-rate curve for the unstratified case at $N=0$ (dashed line) as $Pe$ goes to zero. 
This implies that high thermal diffusivity suppresses the effect of stratification on the inflectional instability. 
{The suppression of the three-dimensional inflectional instability by the thermal diffusion} is also found in fast rotating regime at $N/f=0.1$ (Fig.~\ref{Fig_growth_Pe}b). 
We see that the instability is sustained at smaller $k_{\mathrm{z}}$ for the fast rotation with small $N/f$, while the stabilization with $k_{\mathrm{z}}$ is faster as the thermal diffusivity increases.

%
%                                                One column figure
%----------------------------------------------------------------- 
   \begin{figure}
   \centering
   \includegraphics[width=6cm]{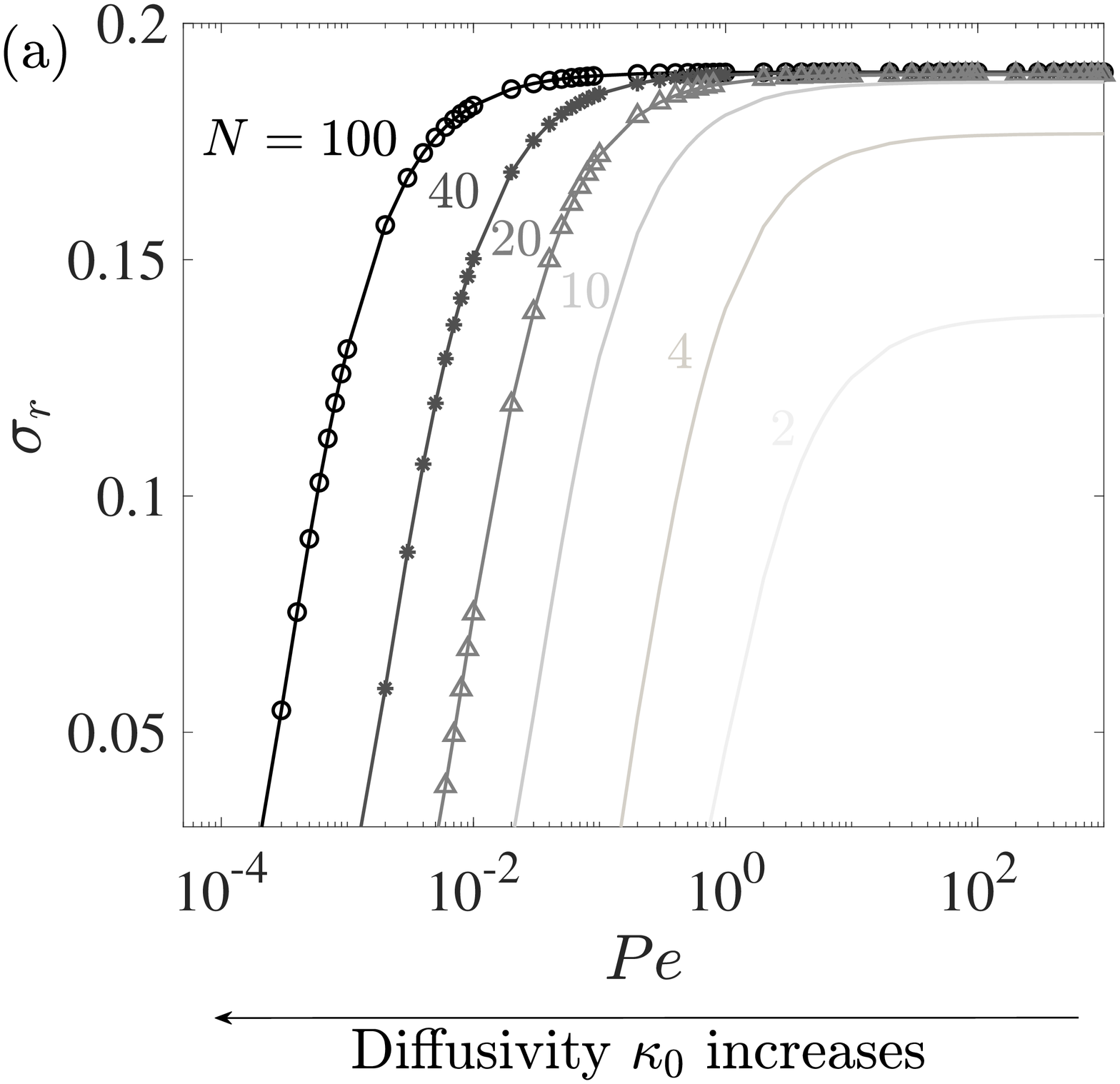}
      \includegraphics[width=6cm]{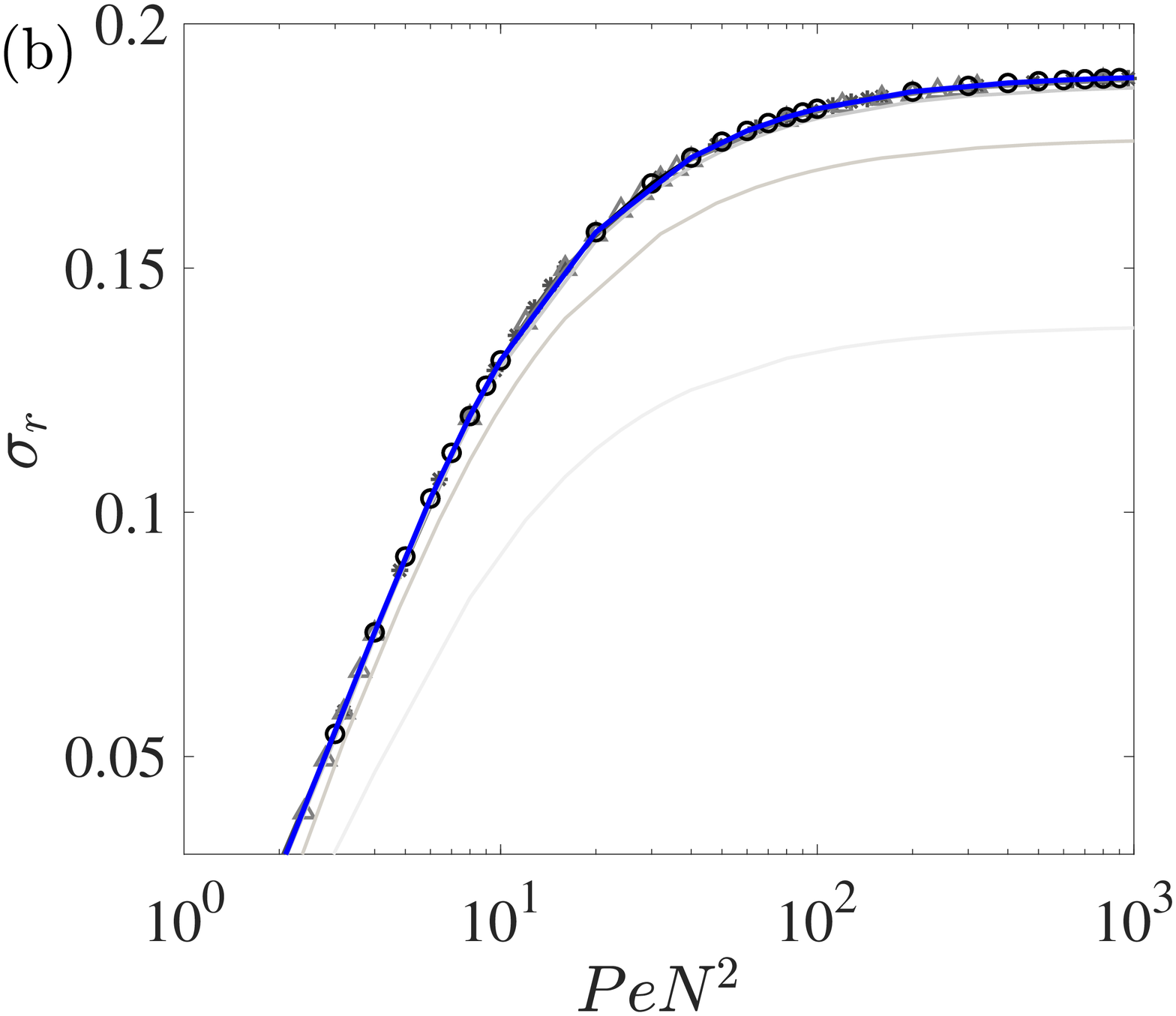}
     \caption{Growth rate of the inflectional instability versus (a) $Pe$ and (b) $PeN^{2}$ for different values of $N$ at $k_{\mathrm{x}}=0.445$, $k_{\mathrm{z}}=1$, and $f=0$.
     Blue line represents the growth rate computed from the eigenvalue problem (\ref{eq:lse_matrix_new}) under the small-$Pe$ approximation.
              }
         \label{Fig_growth_Pe_N}
   \end{figure}
In Fig.~\ref{Fig_growth_Pe_N}, we investigate {the} effects of both $Pe$ and $N$ {on the three-dimensional inflectional instability in the inertially-stable regime at $f=0$, $k_{\rm{x}}=0.445$, and $k_{\rm{z}}=1$.}
We see in Fig.~\ref{Fig_growth_Pe_N}a that the growth rate increases as the stratification becomes strong while the strong thermal diffusion with small $Pe$ suppresses the instability. 
For strong stratification $N\geq10$, it is also observed that the shape of growth-rate curves is similar and they just have different onsets of stabilization at lower P\'eclet numbers (i.e., higher diffusivity) as $N$ increases. 
Due to this resemblance of growth-rate curves for strong stratification, we plot again the growth rate $\sigma_{r}$ versus the rescaled parameter $PeN^{2}$ in Fig.~\ref{Fig_growth_Pe_N}b. We see that the rescaled growth rate curves collapse well for high $N$. This selfsimilarity represented by the rescaled parameter $Pe N^{2}$ is reminiscent of the Richardson-P\'eclet number $RiPe$ in the small-P\'eclet-number approximation used for vertically sheared flows in stratified and thermally-diffusive fluids \citep{Lignieresetal1999,PratLignieres2013}. 

{Following the small-$Pe$ approximation \citep[]{Lignieres1999}, we introduce the rescaled temperature deviation $\hat{\theta}$:
\begin{equation}
\hat{\theta}=\frac{\hat{T}}{Pe}.
\end{equation}
In the limit of small $Pe$, we obtain the set of linear stability equations at leading order:
\begin{equation}
\label{eq:lse_continuity_new}
	\mathrm{i}k_{\mathrm{x}}\hat{u}+\frac{\partial\hat{v}}{\partial y}+\mathrm{i}k_{\mathrm{z}}\hat{w}=0,
\end{equation}
\begin{equation}
\label{eq:lse_x_mom_new}
	\left(\sigma+\mathrm{i}k_{\mathrm{x}} U\right)\hat{u}+\left(U'-f\right)\hat{v}=-\mathrm{i}k_{\mathrm{x}}\hat{p},
\end{equation}
\begin{equation}
\label{eq:lse_y_mom_new}
	\left(\sigma+\mathrm{i}k_{\mathrm{x}} U\right)\hat{v}+f\hat{u}=-\hat{p}_{y},
\end{equation}
\begin{equation}
\label{eq:lse_z_mom_new}
	\left(\sigma+\mathrm{i}k_{\mathrm{x}} U\right)\hat{w}=-\mathrm{i}k_{\mathrm{z}}\hat{p}+PeN^{2}\hat{\theta},
\end{equation}
\begin{equation}
\label{eq:lse_diffusion_new}
\hat{w}=\hat{\nabla}^{2}\hat{\theta}.
\end{equation}
The equations (\ref{eq:lse_continuity_new}-\ref{eq:lse_diffusion_new}) can be simplified into an eigenvalue problem for $\hat{v}$ and $\hat{\theta}$:
\begin{equation}
\label{eq:lse_matrix_new}
\mathcal{C}
\left(
\begin{array}{c}
\hat{v}\\
\hat{\theta}
\end{array}
\right)
=
\sigma
\mathcal{D}
\left(
\begin{array}{c}
\hat{v}\\
\hat{\theta}
\end{array}
\right),
\end{equation}
where
\begin{equation}
\mathcal{C}=\left[
\begin{array}{cc}
\mathcal{C}_{11} & \mathcal{C}_{12}\\
\mathcal{C}_{21} & \mathcal{C}_{22}
\end{array}
\right],~~~~
\mathcal{D}=\left[
\begin{array}{cc}
-\mathrm{i}k_{\rm{z}}\frac{\mathrm{d}}{\mathrm{d}y} & k^{2}\hat{\nabla}^{2}\\
\mathrm{i}\left(k_{\rm{x}}^{2}-\frac{\mathrm{d}^{2}}{\mathrm{d}y^{2}}\right) & k_{\mathrm{z}}\hat{\nabla}^{2}\frac{\mathrm{d}}{\mathrm{d}y}
\end{array}
\right],~
\end{equation}
\begin{equation}
\mathcal{C}_{11}=k_{\mathrm{x}}k_{\rm{z}}\left(U'-f-U\frac{\mathrm{d}}{\mathrm{d}y}\right),
\end{equation}
\begin{equation}
\mathcal{C}_{12}=k_{\mathrm{x}}^{2}PeN^{2}-\mathrm{i}k_{\rm{x}}k^{2}U\hat{\nabla}^{2},
\end{equation}
\begin{equation}
\mathcal{C}_{21}=k_{\rm{x}}\left[U\left(k_{\rm{x}}^{2}-\frac{\mathrm{d}^{2}}{\mathrm{d}y^{2}}\right)+U''\right],
\end{equation}
\begin{equation}
\mathcal{C}_{22}=-\mathrm{i}k_{\rm{x}}k_{\rm{z}}\left[U\hat{\nabla}^{2}\frac{\rm{d}}{\rm{d}y}+(U'-f)\hat{\nabla}^{2}\right].
\end{equation}
We now see that the dependence of the problem on $Pe$ and $N^{2}$ can be studied with the single parameter $PeN^{2}$ in the small-$Pe$ limit. 
As illustrated in Fig.~\ref{Fig_growth_Pe_N}b, the growth rate in the limit of small $Pe$ computed from the eigenvalue problem (\ref{eq:lse_matrix_new}) (blue line in Fig.~\ref{Fig_growth_Pe_N}b) agrees very well with collapsed growth-rate curves plotted against the parameter $PeN^{2}$ for large $N$.
}

\subsection{Effects of $Pe$ on the inertial instability}   
The WKBJ approximation provides explicit dispersion relations for the inertial instability.
They show how the growth rate depends on the stratification in the limit $Pe\rightarrow\infty$ and how the growth rate in stratified fluids becomes equivalent to that in unstratified fluids as $Pe\rightarrow0$.
Nonetheless, it is imperative to investigate whether this argument is valid for any $k_{\mathrm{x}}$ and finite $Pe$ since the WKBJ analysis in this paper is only applied for $k_{\mathrm{x}}=0$ in the two extreme limits: $Pe\rightarrow\infty$ and $Pe\rightarrow0$. In the inertially-unstable regime $f=0.5$ where both the inflectional and inertial instabilities are present (Fig.~\ref{Fig_growth_Pe}c), we see that the growth rate in the range $k_{\mathrm{z}}>0.6$ is increased while the growth rate in the range $k_{\mathrm{z}}<0.6$ is decreased as $Pe$ decreases to zero. 
For both cases, we clearly see that the growth-rate curves in diffusive fluids (solid lines) asymptote to the curve for unstratified case at $N=0$ (dashed line) as $Pe\rightarrow0$. 
{From this asymptotic behavior, we can verify that the growth rate at small wavenumber $k_{\mathrm{z}}<0.6$ corresponds to the inflectional instability that is stabilized as $Pe\rightarrow0$, while the growth rate at large wavenumber $k_{\mathrm{z}}>0.6$ corresponds to the inertial instability that is destabilized as $Pe\rightarrow0$.} 

%
%                                                One column figure
%----------------------------------------------------------------- 
   \begin{figure}
   \centering
      \includegraphics[width=6cm]{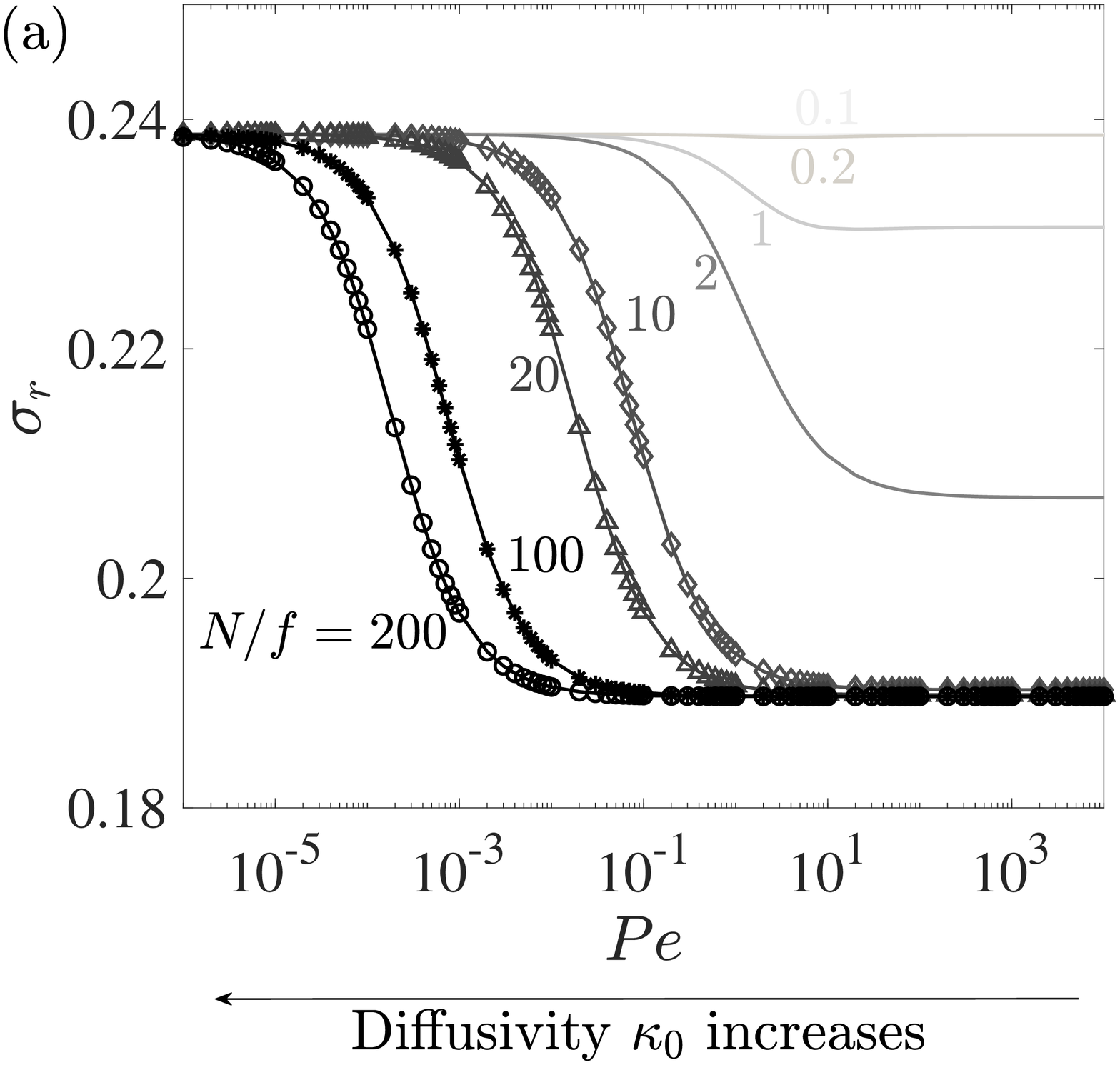}
   \includegraphics[width=6cm]{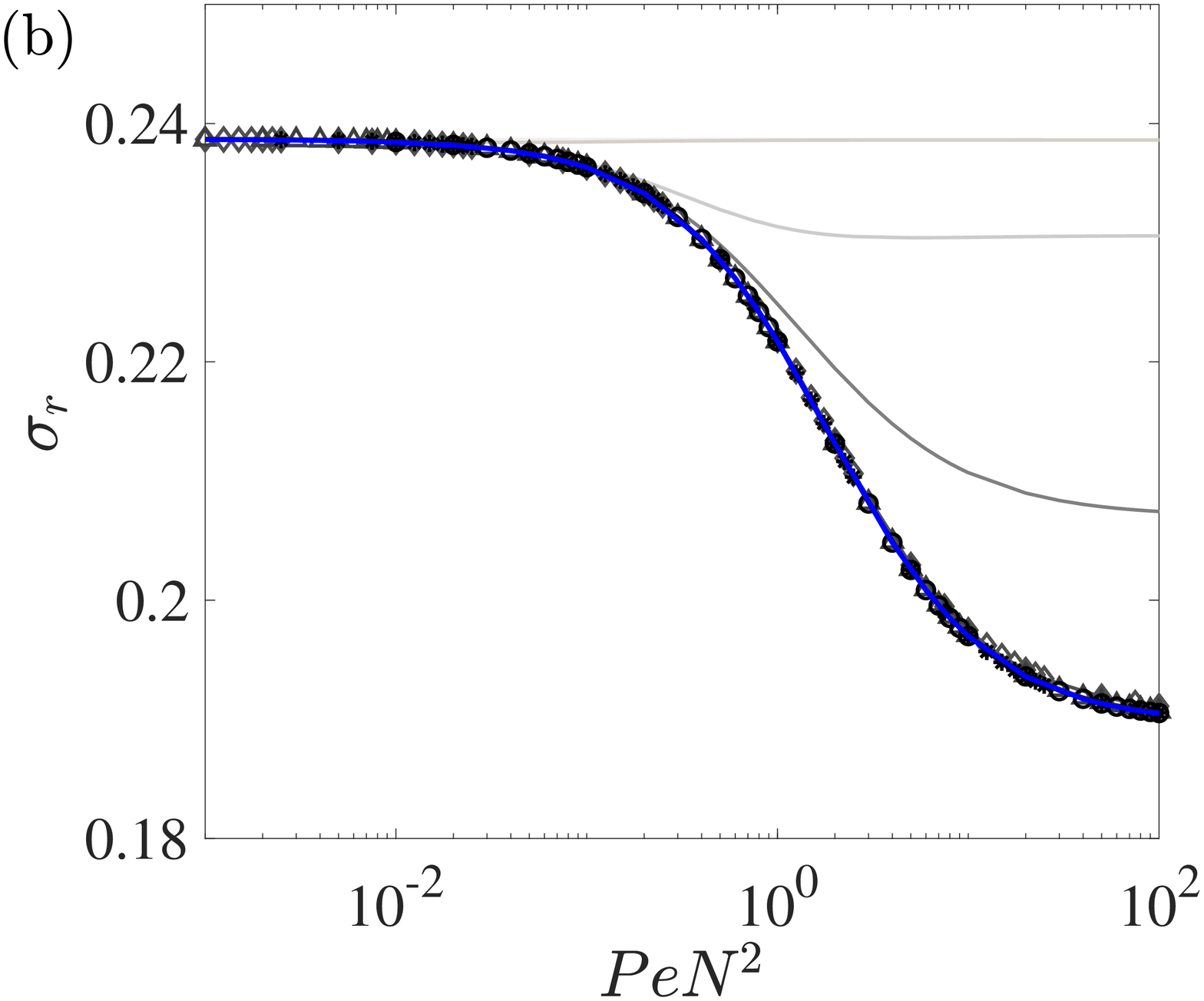}
     \caption{Inertial instability growth rate versus (a) $Pe$ and (b) $PeN^{2}$ for different values of $N/f$ at $k_{\mathrm{x}}=0.445$, $k_{\mathrm{z}}=1$, and $f=0.5$.
     {Blue line denotes the growth rate computed from (\ref{eq:lse_matrix_new}) in the small $Pe$ limit.}
              }
         \label{Fig_growth_Pe_Nf}
   \end{figure}
Picking up the growth rate of the inertial instability at $(k_{\mathrm{x}},k_{\mathrm{z}})=(0.445,1)$, we show in Fig.~\ref{Fig_growth_Pe_Nf} {the} effects of the P\'eclet number $Pe$ on the inertial instability for different values of the ratio $N/f$ at $f=0.5$.
For weakly stratified {cases} with low $N/f<1$, the growth rate remains constant as $\sigma_{r}\simeq0.239$ {for a} wide range of $Pe$. 
On the other hand, the growth rate decreases as $Pe$ increases and asymptotes to $\sigma_{r}\simeq0.190$ for high $N/f$. 
Similar to Fig.~\ref{Fig_growth_Pe_N}a, the shape of growth-rate curves for high ratio $N/f>10$ resemble with different onsets of stabilization as $Pe$ increases. 
As the rescaled parameter $PeN^{2}$ is applied for growth rate curves (Fig.~\ref{Fig_growth_Pe_Nf}b), we see that {the curves for large values of $N/f$ collapse and match with a stability curve computed from the eigenvalue problem (\ref{eq:lse_matrix_new}) under the small-$Pe$ approximation (blue line in Fig.~\ref{Fig_growth_Pe_Nf}b), similarly to the purely inflectional instability case in Fig.~\ref{Fig_growth_Pe_N}b.}

%\begin{figure*}
%   \centering
%   \includegraphics[height=3.85cm]{fig8ab.eps}
%   \includegraphics[height=3.85cm]{fig8cd.eps}
%   %%%\includegraphics{empty.eps}
%   %%%\includegraphics{empty.eps}
 %  \caption{
 %  (a,c) Growth rate $\sigma_{r}$ versus P\'eclet number $Pe$ for different values of $N$ at $k_{\mathrm{x}}=0.445$, $k_{\mathrm{z}}=1$ and (a) $f=0.5$, (b) $f=0$. (b,d) The growth rate curves same as (b,d) but with the rescaled parameter $Pe^{-1}N^{-2}$ for different values of $N$. 
%   }
%              \label{Fig8}%
%    \end{figure*}
%
%
%
%\subsection{Effects of $N/f$}
%
%
%                                                One column figure
%----------------------------------------------------------------- 
   \begin{figure}
   \centering
   \includegraphics[width=6cm]{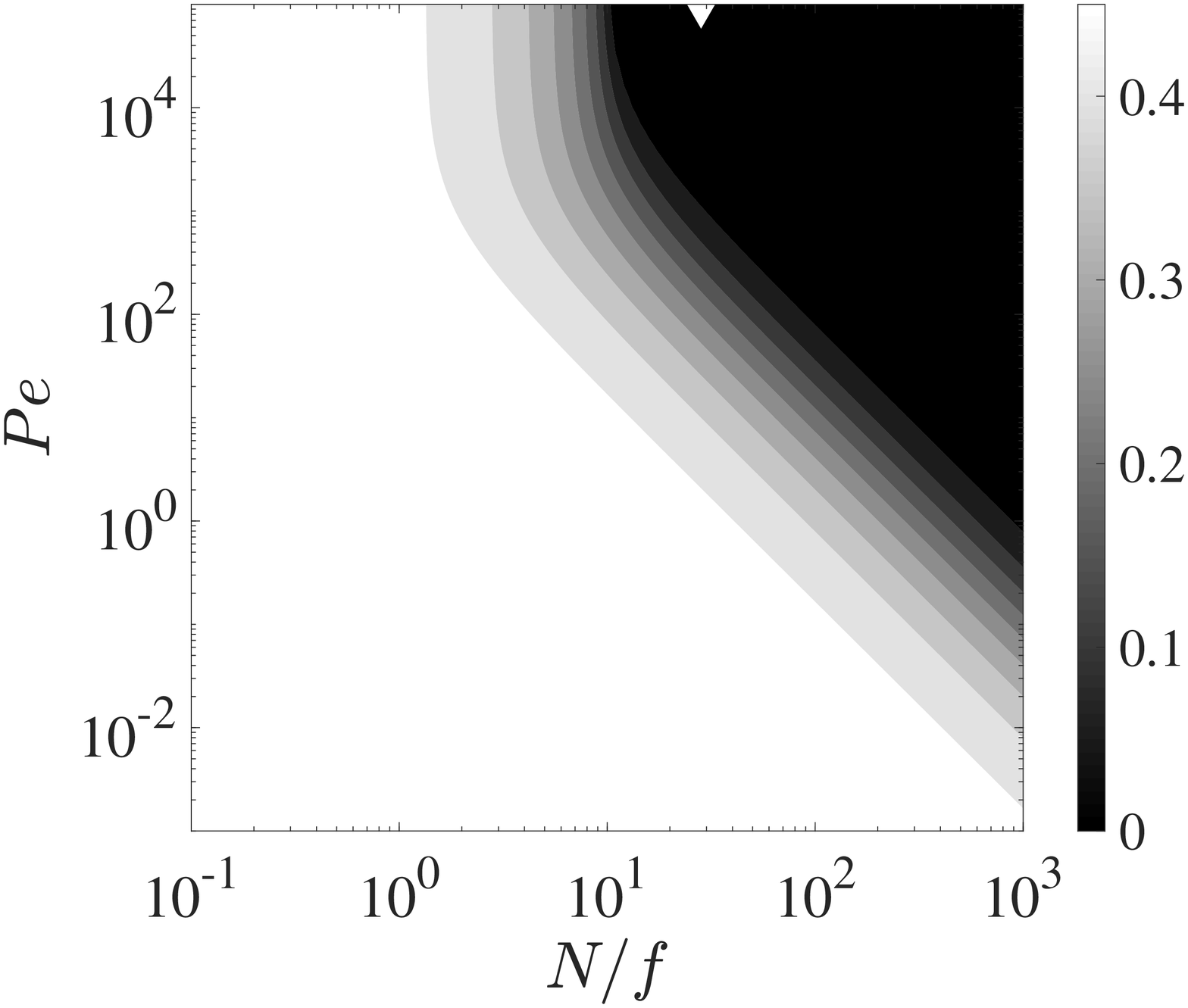}
     \caption{Growth rate contours in the parameter space $(N/f,Pe)$ for $f=0.5$, $k_{\mathrm{x}}=0$, $k_{\mathrm{z}}=20$. The white triangle on the axis denotes the critical $N/f$ from (\ref{eq:critical_Nf}) as $Pe\rightarrow\infty$.  
%   \includegraphics[width=6.5cm]{fig7a.eps}
%   \includegraphics[width=6.5cm]{fig7b.eps}
%      \caption{Growth rate $\sigma_{r}$ versus $f/N$ at $f=0.5$ and (a) $(k_{\mathrm{x}},k_{\mathrm{z}})=(0.445,0.2)$, (b) $(k_{\mathrm{x}},k_{\mathrm{z}})=(0,10)$ for different P\'eclet numbers $Pe$: $Pe=\infty$ (black solid lines), $Pe=\{100,10,1,0.1,0.01,0.001,0.0001\}$ (gray solid lines). Dashed lines indicate the growth rate for unstratified case $N=0$. Arrows indicate the direction of decreasing $Pe$. 
              }
         \label{Fig_contours_Nf_Pe}
   \end{figure}
%-----------------------------------------------------------------
Fig.~\ref{Fig_contours_Nf_Pe} shows contours of the growth rate $\sigma_{r}$ in the parameter space $(N/f,Pe)$ at $k_{\mathrm{x}}=0$, $k_{\mathrm{z}}=20$, and $f=0.5$, a typical parameter set for the inertial instability. 
We see that for a fixed $N/f$, the inertial instability destabilizes as the thermal diffusivity increases (i.e., as $Pe\rightarrow 0$), similarly to what is observed in Fig.~\ref{Fig_growth_Pe_Nf}. 
{For a fixed $Pe$, the growth rate decreases as $N/f$ increases (i.e., the stratification stabilizes the system at a fixed rotation).} 
{This can be mathematically predicted from the case for nondiffusive fluids as $Pe\rightarrow\infty$} by the asymptotic dispersion relation (\ref{eq:WKBJ_2ODE_dispersion_1st}) as the term $\sigma_{1}$ increases with $N$ (i.e., the growth rate decreases with $N$). 
We can further derive from (\ref{eq:WKBJ_2ODE_dispersion_1st}) the critical value of the ratio $N/f$ in the limit $Pe\rightarrow\infty$ where the growth rate $\sigma$ becomes zero:
\begin{equation}
\label{eq:critical_Nf}
	\frac{N}{f}\biggr|_{\mathrm{crit}}=\sqrt{\frac{(1-f)^{2}k_{\mathrm{z}}^{2}}{(m-1/2)^{2}f}-\left(\frac{1-f}{f}\right)}.
\end{equation}
{We see in Fig.~\ref{Fig_contours_Nf_Pe} that the asymptotic prediction (\ref{eq:critical_Nf}) for the critical value of $N/f$ lies in the stable regime obtained from numerical results. 
Also, it is notable from the equation (\ref{eq:critical_Nf})} that for a fixed $f$, the critical ratio $N/f$ increases with the vertical wavenumber $k_{\mathrm{z}}$. This implies that the characteristic vertical length scale $\lambda_{z}=2\pi/k_{\mathrm{z}}$ decreases as the ratio $N/f$ increases, {as observed in other stratified-rotating flows \citep[see e.g.,][]{Caton2000}}.

\section{Conclusion}
\begin{table*}[!t]
\caption{Summary table for the growth rate and its variation with parameters $Pe$, $N$, $f$, and $(k_{\mathrm{x}},k_{\mathrm{z}})$ for the {three-dimensional inflectional instability (i.e., $k_{\rm{z}}\geq0$) and the inertial instability.}}
\label{table:1}
\centering
\begin{tabular}{l | c | c | c | c}
\hline
\hline
Instability type & $Pe\downarrow$ $(\kappa_{0}\uparrow)$ & $N\uparrow$ & Values of $\max(\sigma_{r})$ & Corresponding wavenumbers\\
\hline
Inflectional & $\downarrow$ & $\uparrow$ & 0.1897 (independent of $f$) & $k_{\rm{x}}=0.445,~~k_{\rm{z}}\rightarrow0$\\
Inertial  & $\uparrow$ & $\downarrow$ & $\sqrt{f(1-f)}$ (maximum at $f=0.5$) & $k_{\rm{x}}=0,~~k_{\rm{z}}\rightarrow\infty$\\
\hline
\end{tabular}
\end{table*}
This paper investigates {the} instabilities of horizontal shear flow in stably-stratified, rotating, and thermally-diffusive fluids corresponding to stellar radiative regions. 
On the one hand, the inflectional shear instability always exists for the horizontal shear flow in a hyperbolic tangent profile whose maximum growth rate $\sigma_{\rm max}=0.1897$ is attained at $k_{\mathrm{x}}=0.445$ and $k_{\mathrm{z}}=0$ independently of the stratification, rotation, and thermal diffusion. 
{For the three-dimensional inflectional instability for nonzero vertical wavenumber $k_{\rm{z}}>0$, the stable stratification that is inhibited by thermal diffusion has a destabilizing action.}
This is the opposite of the case of the vertical shear instability, which is inhibited by the stable stratification but favored by thermal diffusion that again diminishes its effects \cite[e.g.,][]{Zahn1983,Zahn1992}. 
On the other hand, the inertial instability is present in the range of $0<f<1$ and its maximum growth rate $\sigma_{\rm max}=\sqrt{f(1-f)}$ is reached as $k_{\mathrm{z}}\rightarrow\infty$ in the inviscid limit for both nondiffusive and high-diffusivity fluids. The analysis on the inertial instability for the case $k_{\mathrm{x}}=0$ and large $k_{\mathrm{z}}$ has been elaborated further {through} the WKBJ approximation in two limits: $Pe\rightarrow\infty$ and $Pe\rightarrow0$ (i.e., for low and high thermal diffusivity, respectively), and explicit expressions for asymptotic dispersion relations are provided in both limits. 
For $Pe\rightarrow\infty$, the growth rate decreases with $N$ thus the stratification stabilizes the inertial instability, but the maximum growth rate $\sigma_{\rm max}$ at infinite $k_{\mathrm{z}}$ remains as $\sqrt{f(1-f)}$ independently of $N$. 
In the limit $Pe\rightarrow0$, the growth rate is no longer dependent on the stratification and becomes identical to {the inertial instability growth rate} for the unstratified case at $N=0$. 
Detailed numerical studies confirm in the general parameter space $(k_{\mathrm{x}},k_{\mathrm{z}})$ that both the inflectional and inertial instabilities in thermally-diffusive fluids asymptote to those of the unstratified case as the thermal diffusivity increases (i.e., $Pe\rightarrow0$).
{The} selfsimilarity of the growth rate in stratified-rotating fluids is also found such that the instabilities depend on the parameter $PeN^{2}$ for small $Pe$ and strong stratification $N$. 
As a summary, we describe in Table~\ref{table:1} the growth rate and its variation with parameters $Pe$, $N$, $f$, $k_{\mathrm{x}}$ and $k_{\mathrm{z}}$. 

The present work brings to light two horizontal shear instabilities that probably occur in stellar radiative zones but that had not been considered thoroughly before in stellar astrophysics, especially for the case of inertial instability on a local $f$-plane with the effects of thermal diffusion. 
The particularity of the stellar regime is that the high thermal diffusivity {can} weaken the stabilizing effect of the stratification for the inertial instability thus to enhance its development, while the three-dimensional inflectional instability is suppressed by the high thermal diffusivity and the inflectional instability mode becomes dominantly two-dimensional.
We first investigated the linear instabilities in the polar regions but their nonlinear saturation and the resulting anisotropic turbulent transport of angular momentum and chemicals in both the horizontal and vertical directions has to be quantified. 
To derive the associated prescriptions for stellar evolution models, it is imperative to study the effects of the complete Coriolis acceleration on the instabilities at any co-latitude using the so-called nontraditional $f$-plane approximation \citep[]{Gerkema2008}. This will be done in the next article of the series.

\begin{acknowledgements}
The authors acknowledge support from the European Research Council through ERC grant SPIRE 647383 and from GOLF and PLATO CNES grants at the Department of Astrophysics of CEA.
We appreciate the referee's helpful comments and suggestions to improve the manuscript.
And we also thank Dr. K. Augustson for his careful reading and comments on the manuscript. 
\end{acknowledgements}

%\begin{appendix} %First appendix
%\section{Matrix coefficients}
%\label{appendix:matrix}
%The matrices $\mathcal{A}$ and $\mathcal{B}$ in (\ref{eq:lse_matrix}) are expressed as
%\begin{equation}
%\label{eq:appendix_matrixA}
%	\mathcal{A}=
%	\left[
%	\begin{array}{ccc}
%	\mathcal{A}_{11} & \mathcal{A}_{12} & 0\\
%	\mathrm{i}k^{2}f & \mathcal{A}_{22} & N^{2}k_{\mathrm{z}}\frac{\mathrm{d}}{\mathrm{d}y}\\
%	\mathrm{i}k_{\mathrm{x}} & \frac{\mathrm{d}}{\mathrm{d}y} & k_{\mathrm{x}} k_{\mathrm{z}} U+\frac{\mathrm{i}k_{\mathrm{z}}}{Pe}\hat{\nabla}^{2}
%	\end{array}
%	\right],
%\end{equation}
%\begin{equation}
%\label{eq:appendix_matrixB}
%	\mathcal{B}=
%	\left[
%	\begin{array}{ccc}
%	-\frac{\mathrm{d}}{\mathrm{d}y} & \mathrm{i}k_{\mathrm{x}} & 0\\
%	0 & \mathrm{i}\hat{\nabla}^{2} & 0\\
%	0 & 0 & \mathrm{i}k_{\mathrm{z}}
%	\end{array}
%	\right],
%\end{equation}
%where
%\begin{eqnarray}
%\mathcal{A}_{11}&=&\mathrm{i}k_{\mathrm{x}}\left(U'+U\frac{\mathrm{d}}{\mathrm{d}y}\right)-\mathrm{i}k_{\mathrm{x}} f,\nonumber\\
%\mathcal{A}_{12}&=&k_{\mathrm{x}}^{2}U+\left(U'-f\right)\frac{\mathrm{d}}{\mathrm{d}y}+U'',\nonumber\\
%\mathcal{A}_{22}&=&k_{\mathrm{x}} U\hat{\nabla}^{2}+fk_{\mathrm{x}}\frac{\mathrm{d}}{\mathrm{d}y}-k_{\mathrm{x}} U''.
%\end{eqnarray}
%\end{appendix}

\bibliographystyle{aa} % style aa.bst
\bibliography{aa} % your references Yourfile.bib

\end{document}